\documentclass[10pt,twocolumn,twoside]{IEEEtran}

\usepackage{graphicx}
\usepackage{subfigure}
\usepackage{setspace}
\usepackage{multicol}
\usepackage{multirow}
\usepackage{amsmath}
\usepackage{bm}
\usepackage{cite}
\usepackage{amssymb}
\usepackage{gensymb}
\usepackage{amsfonts}
\usepackage{mathrsfs}
\usepackage{amsmath}
\usepackage{algorithm}
\usepackage{algorithmic}
\usepackage{amsthm}

\usepackage{tabularx}
\usepackage{color}
\usepackage{balance}
\usepackage{mathrsfs}
\usepackage{setspace}

\usepackage{array}
\newcommand{\PreserveBackslash}[1]{\let\temp=\\#1\let\\=\temp}
\newcolumntype{C}[1]{>{\PreserveBackslash\centering}p{#1}}
\usepackage[flushleft]{threeparttable}
\usepackage{setspace}

\begin{document}

\bibliographystyle{IEEEtran} 

\title{Delay-Phase Precoding for Wideband THz Massive MIMO}

\author{
	Linglong Dai, {\it Fellow, IEEE}, Jingbo Tan, {\it Student Member, IEEE}, Zhi Chen, {\it Senior Member, IEEE}, and H. Vincent Poor, {\it Life Fellow, IEEE}
	\vspace{-5mm}
	\thanks{	
	A part of this paper was presented in the IEEE
	Global Communications Conference (GLOBECOM'19)\cite{Ref:DPP2020}. 	
			
	L. Dai and J. Tan are with the Beijing National Research Center for Information Science and Technology (BNRist) as well as the Department of
	Electronic Engineering, Tsinghua University, Beijing 100084, China (E-mails: daill@tsinghua.edu.cn; tanjb17@mails.tsinghua.edu.cn). Z. Chen is with the National Key Laboratory of Science and Technology on Communications, University of Electronic Science and Technology of China, Chengdu 611731, China (e-mail: chenzhi@uestc.edu.cn). H. V. Poor is with the Department of Electrical and Computer Engineering, Princeton University, Princeton, NJ 08544, USA (E-mail: poor@princeton.edu).
	
	This work was supported in part by the National Natural Science Foundation of China (Grant No. 62031019), in part by the National Key Research and Development Program of China (Grant No. 2020YFB1805005), and in part by the European Commission through the H2020-MSCA-ITN META WIRELESS Research Project under Grant 956256.
	}
	
}

\maketitle
\IEEEpeerreviewmaketitle
\begin{abstract}
Benefiting from tens of GHz of bandwidth, terahertz (THz) communication has become a promising technology for future 6G network. To deal with the serious propagation loss of THz signals, massive multiple-input multiple-output (MIMO) with hybrid precoding is utilized to generate directional beams with high array gains. However, the standard hybrid precoding architecture based on frequency-independent phase-shifters cannot cope with the beam split effect in THz massive MIMO caused by the large bandwidth and the large number of antennas, where the beams split into different physical directions at different frequencies. The beam split effect will result in a serious array gain loss across the entire bandwidth, which has not been well investigated in THz massive MIMO. In this paper, we first quantify the seriousness of the beam split effect in THz massive MIMO by analyzing the array gain loss it causes. Then, we propose a new precoding architecture called delay-phase precoding (DPP) to mitigate this effect. Specifically, the proposed DPP introduces a time delay network composed of a small number of time delay elements between radio-frequency chains and phase-shifters in the standard hybrid precoding architecture. Unlike \emph{frequency-independent} phase shifts, the time delay network introduced in the DPP can realize \emph{frequency-dependent} phase shifts, which can be designed to generate frequency-dependent beams towards the target physical direction across the entire bandwidth. Due to the joint control of delay and phase, the proposed DPP can alleviate the array gain loss caused by the beam split effect. Furthermore, we propose a hardware structure by using true-time-delayers to realize frequency-dependent phase shifts for realizing the concept of DPP. A corresponding precoding algorithm is proposed to realize the precoding design. Theoretical analysis and simulations show that the proposed DPP can mitigate the beam split effect and achieve near-optimal rate with higher energy efficiency.
\end{abstract}


\section{Introduction}
The future 6G wireless network is expected to realize Tbps single-user data rate to support emerging ultra-high-speed applications, such as mobile holograms, immersive virtual reality, and digital twins\cite{Ref:To6G2020}. To realize such a rapid growth of the data traffic in future 6G wireless network, terahertz (THz) communication is considered to be a promising technology due to the tenfold increase in bandwidth it can provide \cite{Ref:Thz100GHz2019,Ref:Be5GThz2018,Ref:SurThz2019,Ref:THzdistance2018,Ref:TeraBand2014,Ref:Tera2011}. Compared with the typical bandwidth of several GHz in the millimeter-wave (mmWave) band ($30$-$100$ GHz) for 5G, the THz band ($0.1$-$10$ THz) for 6G is capable of providing at least $10$ GHz bandwidth or even much larger\cite{Ref:mmWaveMIMO2016,Ref:TeraBand2014}. However, THz signals suffer from severe propagation attenuation due to the very high carrier frequencies, which is a major obstacle to practical THz communications\cite{Ref:Thz100GHz2019}. Massive multiple-input multiple-output (MIMO), which utilizes a large antenna array to generate directional beams with high array gains, can be used to compensate for such severe signal attenuation in THz communications. Consequently, THz massive MIMO is very promising for future 6G wireless communications\cite{Ref:AoATera2018,Ref:BDMATera2017,Ref:TrackTera2017}. Similar to mmWave massive MIMO, hybrid precoding has been recently considered for THz massive MIMO to relieve the substantial power consumption of THz radio-frequency (RF) chains\cite{Ref:HybridMili2015}. The key idea of hybrid precoding is to decompose the high-dimensional fully-digital precoder into a high-dimensional analog beamformer realized by phase-shifters (PSs) and a low-dimensional digital precoder\cite{Ref:LowRF2018}. Thus, a significantly reduced number of RF chains can be used. Thanks to the sparsity of THz channels\cite{Ref:THzChannel2019}, a small number of RF chains are still sufficient to fully achieve the multiplexing gain in THz massive MIMO systems\cite{Ref:LowRF2018,Ref:SpatiallyPre2014}.
\vspace{-2mm}
\subsection{Prior works}
In the conventional hybrid precoding architecture, the analog beamformer will generate directional beams aligned with the physical directions of channel path components to realize the full array gain\cite{Ref:SpatiallyPre2014}. Such an analog beamformer works well for narrowband systems. However, for wideband 5G mmWave massive MIMO systems, the beams at different subcarrier frequencies will point to different physical directions due to use of the  \emph{frequency-independent} PSs\cite{Ref:SpatiallyPre2014}, which results in an array gain loss. To deal with the array gain loss incurred by this effect, called beam squint\cite{Ref:WidebandEff2019}, several methods have been proposed for mmWave massive MIMO systems \cite{Ref:SubWideHy2017,Ref:WideMIMOHy2018,Ref:WideCode2016,Ref:WideCode2019}. Specifically, the hybrid precoding problem in orthogonal frequency division multiplexing (OFDM) based wideband massive MIMO systems was formulated in \cite{Ref:SubWideHy2017}, where a near-optimal closed-form solution was developed. Aiming at improving the performance of hybrid precoding, an alternating optimization algorithm was proposed in \cite{Ref:WideMIMOHy2018}, which iteratively optimized the analog beamformer and digital precoder to achieve near-optimal rate performance across the entire bandwidth. In addition, codebooks containing beams with wide beamwidths were designed to reduce the array gain loss caused by the beam squint effect in \cite{Ref:WideCode2016,Ref:WideCode2019}. Specifically, the wide beams were designed in \cite{Ref:WideCode2016} by maximizing the minimum array gain achieved at all subcarriers, while a semidefinite relaxation method was utilized in \cite{Ref:WideCode2019} to maximize the total array gain across the entire bandwidth. These methods \cite{Ref:SubWideHy2017,Ref:WideMIMOHy2018,Ref:WideCode2016,Ref:WideCode2019} are effective for improving the achievable rate performance, as the beams only slightly squint and the array gain loss is not serious in wideband mmWave massive MIMO systems. 

However, the methods of \cite{Ref:SubWideHy2017,Ref:WideMIMOHy2018,Ref:WideCode2016,Ref:WideCode2019} are not valid for wideband THz massive MIMO systems. Due to the much wider bandwidth of THz signals and the much larger number of antennas to be used, beams at different subcarriers will split into totally separated physical directions. This effect, called \emph{beam split} in this paper, which is a key difference between mmWave and THz massive MIMO systems, brings a new fundamental challenge for THz communications. Specifically, unlike the situation under the beam squint effect where the beams can still cover the user across the entire bandwidth, the beams generated by frequency-independent PSs can only be aligned with the target user over a small portion of all subcarriers around the central frequency due to the beam split effect. This indicates that only the beams around the central frequency can achieve high array gain, while the beams at other subcarriers suffer from a serious array gain loss. Therefore, the beam split effect will result in a severe achievable rate degradation, and counteract the achievable rate gain benefiting from the bandwidth increase in THz massive MIMO systems. Using true-time-delayers to generate frequency-dependent phase shifts is a solution to solve the beam split. For example, deploying one true-time-delayer to control each antenna element can realize wideband beamforming \cite{Ref:TTD2008,Ref:TTD2019} and beam training \cite{Ref:TTDTrain2020}. In order to reduce the required range of time delays, a hybrid true-time-delayer architecture was proposed in \cite{Ref:HybridTTD2021} for fast beam training, where the time delays were realized separately by analog time-delayers with the equal number of antennas and a small number of digital time-delayers. However, since delayers with the equal number of antennas are utilized, the existing works \cite{Ref:TTD2008,Ref:TTD2019,Ref:TTDTrain2020,Ref:HybridTTD2021} may bring high power consumption due to the large number of antennas, especially for precoding where multiple data streams are transmitted. On the other hand, these true-time-delayers based architectures \cite{Ref:TTD2008,Ref:TTD2019,Ref:TTDTrain2020,Ref:HybridTTD2021} mainly consider beamformer design with a single data stream, while the precoding design with multiple data streams has not been considered. In conclusion, to our best knowledge, the beam split effect has not been characterized and investigated in THz massive MIMO systems, and there are no practical solutions for precoding design to overcome this fundamental challenge.
\vspace{-2mm}
\subsection{Our contributions}
In this paper, we first analyze the performance loss caused by the beam split effect, and then we propose the delay-phase precoding (DPP) architecture to mitigate the beam split effect in THz massive MIMO systems. The contributions of this paper can be summarized as follows.
\begin{itemize}
	\item We first reveal and quantify the beam split effect, i.e., the THz rainbow, in wideband THz massive MIMO systems. The relationship between the array gain loss and the system parameters is analyzed. Based on this analysis, we define a metric called beam split ratio to evaluate the degree of the beam split effect, which clearly shows how serious the beam split effect is caused by the large bandwidth and large number of antennas in wideband THz massive MIMO systems.
	\item We propose a new precoding architecture, namely DPP, to mitigate the beam split effect. In the proposed DPP, a time delay (TD) network composed of a small number of TD elements, is introduced between the RF chains and frequency-independent PSs in the conventional hybrid precoding architecture. The PSs are still used to generate beams aligned with the target physical direction, while the time delays in the TD network are designed to make the beams aligned with the target physical directions across the entire bandwidth. In this way, the DPP architecture can convert \emph{frequency-independent} phase controlled beamforming into \emph{frequency-dependent} delay-phase controlled beamforming, which can significantly alleviate the array gain loss caused by the beam split effect. 
	\item A hardware structure called true-time-delayers based DPP (TTD-DPP) is proposed to realize the concept of DPP, where the TD network is realized by a small number of TTDs between RF chains and the PS network. A corresponding precoding algorithm is proposed to realize precoder design supporting multiple data streams. The analysis and simulation results illustrate that the proposed TTD-DPP structure is able to achieve the near-optimal rate with higher energy efficiency\footnote{Simulation codes are provided to reproduce the results in this paper: http://oa.ee.tsinghua.edu.cn/dailinglong/publications/publications.html.}.
\end{itemize}
\vspace{-2mm}
\subsection{Organization and notation}
\emph{Organization:} The remainder of the paper is organized as follows. Section \ref{Sys} introduces the system model of wideband THz massive MIMO. In Section \ref{Squint}, we analyze the beam split effect, and propose the DPP architecture. Then, a  TTDs based hardware structure is proposed to realize DPP in Section \ref{SPOB}. Section \ref{Sim} and \ref{Con} provide simulation results and conclusions, respectively. 

\emph{Notation:} Lower-case and upper-case boldface letters represent
vectors and matrices, respectively; $(\cdot)^{T}$, $(\cdot)^{H}$, $\|\cdot\|_\mathrm{F}$, and $\|\cdot\|_{k}$ denote the transpose, conjugate transpose, Frobenius norm, and $k$-norm of a matrix, respectively; $\mathbf{H}_{[i,j]}$ denotes the element of matrix $\mathbf{H}$ at the $i$-th row and the $j$-th column; $\mathbb{E}(\cdot)$ denotes the expectation; $|\cdot|$ denotes
the absolute value; $\mathbf{I}_{N}$ represents $N\times N$ identity matrix; $\mathrm{blkdiag}(\mathbf{A})$ denotes a block diagonal matrix where each column of $\mathbf{A}$ represents the diagonal blocks of the matrix $\mathrm{blkdiag}(\mathbf{A})$; $\mathcal{CN}(\mathbf{\mu},\mathbf{\Sigma})$ denotes the Gaussian distribution with mean $\mathbf{\mu}$ and covariance $\mathbf{\Sigma}$; $\mathcal{U}(a,b)$ represents the uniform distribution between $a$ and $b$.

\begin{figure}
	\centering
	\includegraphics[width=0.47\textwidth]{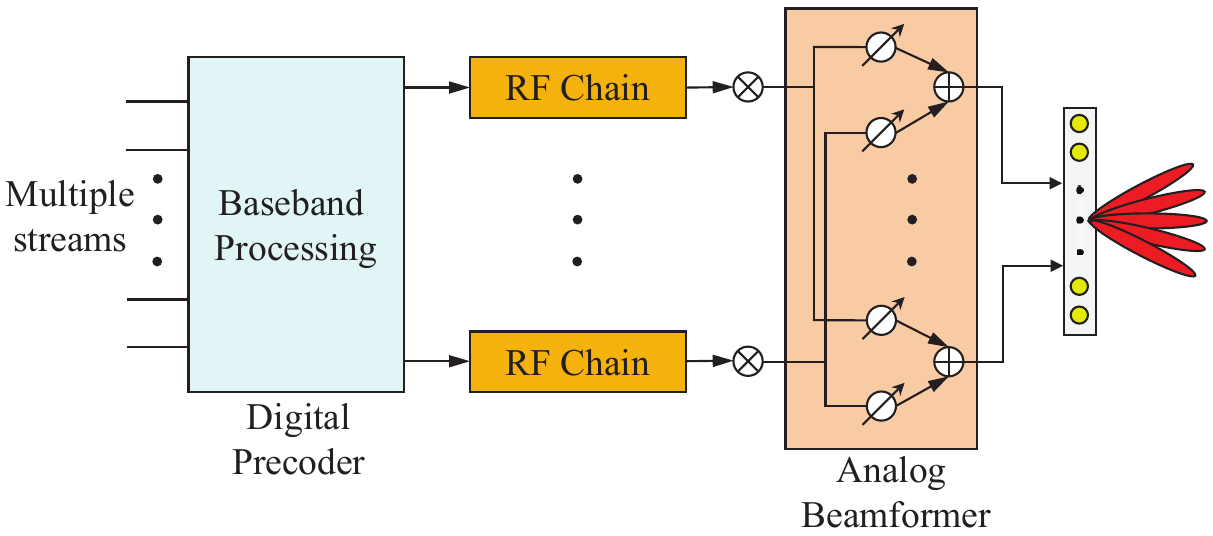}
	\caption{The conventional hybrid precoding architecture\cite{Ref:SpatiallyPre2014}.}
\end{figure}
\vspace{-2mm}
\section{System Model of THz Massive MIMO}\label{Sys}
We first consider a THz massive MIMO system with  conventional hybrid precoding as shown in Fig. 1. The base station (BS) employs $N_\mathrm{RF}$ RF chains and an $N_\mathrm{t}$-antenna uniform linear array (ULA)\footnote{In this paper, we consider the ULA for simplicity, but the analysis of the beam split effect and the correspondingly proposed DPP architecture can be easily extended to the uniform planar array (UPA)\cite{Ref:FundWC2005}, which has a similar channel form as ULA.}. An $N_\mathrm{r}$-antenna user is served and $N_\mathrm{s}$ data streams are transmitted simultaneously by the BS (usually we have $N_\mathrm{s}=N_\mathrm{r}\leq N_\mathrm{RF}\ll N_\mathrm{t}$). The widely used OFDM with $M$ subcarriers is considered. The downlink received signal $\mathbf{y}_{m}\in\mathcal{C}^{N_{r}\times 1}$ at the $m$-th subcarrier can be expressed as\cite{Ref:SpatiallyPre2014}
\begin{equation}\label{1}
\mathbf{y}_{m}=\sqrt{\rho}\mathbf{H}_{m}^{H}\mathbf{A}\mathbf{D}_{m}\mathbf{s}_{m}+\mathbf{n}_{m},
\end{equation}
where $\mathbf{H}_{m}\in\mathcal{C}^{N_\mathrm{t}\times N_\mathrm{r}}$ is the frequency-domain channel at the $m$-th subcarrier, $\mathbf{A}\in\mathcal{C}^{N_\mathrm{t}\times N_\mathrm{RF}}$ with constraint $|\mathbf{A}_{[i,j]}|=\frac{1}{\sqrt{N_\mathrm{t}}}$ is the frequency-independent analog beamformer identical over all $M$ subcarriers which is realized by PSs\cite{Ref:OverMilliMIMO2016}, $\mathbf{D}_{m}\in\mathcal{C}^{N_\mathrm{RF}\times N_\mathrm{s}}$ is the frequency-dependent digital precoder at the $m$-th subcarrier satisfying transmission power constraint $\|\mathbf{A}\mathbf{D}_{m}\|_\mathrm{F}^{2}=N_\mathrm{s}$, $\mathbf{s}_{m}\in\mathcal{C}^{N_\mathrm{s}\times 1}$ is the transmitted signal at the $m$-th subcarrier with $\mathbb{E}(\mathbf{s}_{m}\mathbf{s}_{m}^{H})=\frac{1}{N_\mathrm{s}}\mathbf{I}_{N_\mathrm{s}}$, $\rho$ is the average received power, and $\mathbf{n}_{m}\in \mathcal{C}^{N_\mathrm{s}\times 1}$ is the additive white Gaussian noise at the $m$-th subcarrier following Gaussian distribution $\mathcal{CN}(0,\sigma^{2}\mathbf{I}_{N_\mathrm{s}})$ with $\sigma^{2}$ being the noise power.

In this paper, we consider the widely used wideband ray-based channel model\cite{Ref:AoATera2018} for THz communications. We denote $f_\mathrm{c}$ as the central frequency and $B$ as the bandwidth. Then, the time-domain channel $h_{n_\mathrm{t},n_\mathrm{r}}$ between the $n_\mathrm{t}$-th antenna of the BS and the $n_\mathrm{r}$-th antenna of the user with $n_\mathrm{t}\in1,2,\cdots,N_\mathrm{t}$ and $n_\mathrm{r}\in1,2,\cdots,N_\mathrm{r}$ can be denoted as
\begin{equation}\label{2}
h_{n_\mathrm{t},n_\mathrm{r}}=\sum_{l=1}^{L}g_{l}\delta(t-\tau_{l}-(n_\mathrm{t}-1)\frac{d}{c}\sin{\tilde{\theta}_{l}}-(n_\mathrm{r}-1)\frac{d}{c}\sin{\tilde{\phi}_{l}}),
\end{equation}
where $L$ denotes the number of resolvable paths, $g_{l}$ and $\tau_{l}$ represent the path gain and path delay of the $l$-th path, $\tilde{\theta}_{l},\tilde{\phi}_{l}\in[-\pi/2,\pi/2]$ are the frequency-independent physical directions of the $l$-th path at the BS side and the user side respectively, $d$ is the antenna spacing usually set as $d=\frac{\lambda_\mathrm{c}}{2}=\frac{c}{2f_\mathrm{c}}$ with $\lambda_\mathrm{c}$ denoting the wavelength at the central frequency $f_\mathrm{c}$ and cannot be changed after the antenna array has been fabricated, and $c$ denotes the light speed. In (\ref{2}), $(n_\mathrm{t}-1)\frac{d}{c}\sin{\tilde{\theta}_{l}}$ and $(n_\mathrm{r}-1)\frac{d}{c}\sin{\tilde{\phi}_{l}}$ denote the time delays caused by the physical directions of the $l$-th path at the $n_\mathrm{t}$-th antenna of the BS and that at the $n_\mathrm{r}$-th antenna of the user, respectively.

For the $m$-th subcarrier with frequency $f_{m}=f_\mathrm{c}+\frac{B(2m-1-M)}{2M}$, the frequency-domain channel $\mathbf{H}_{m}$ can be denoted by the discrete Fourier transform (DFT) of the time-domain channel as
\begin{equation}\label{3}
\mathbf{H}_{m}=\sum_{l=1}^{L}g_{l}e^{-j2\pi\tau_{l}f_{m}}\mathbf{f}_\mathrm{t}(\bar{\theta}_{l,m})\mathbf{f}_\mathrm{r}(\bar{\phi}_{l,m})^{H},
\end{equation}
where $\mathbf{f}_\mathrm{t}(\bar{\theta}_{l,m})$ and  $\mathbf{f}_\mathrm{r}(\bar{\phi}_{l,m})$ are the array responses at the BS side and the user side. Taking the array response at the BS as an example, $\mathbf{f}_\mathrm{t}(\bar{\theta}_{l,m})$ can be presented by setting $d=\frac{c}{2f_\mathrm{c}}$ as 
\begin{equation}\label{4}
\mathbf{f}_\mathrm{t}(\bar{\theta}_{l,m})=\frac{1}{\sqrt{N_\mathrm{t}}}\left[1,e^{j\pi\bar{\theta}_{l,m}},e^{j\pi 2\bar{\theta}_{l,m}},\cdots,e^{j\pi(N_\mathrm{t}-1)\bar{\theta}_{l,m}}\right]^{H},
\end{equation}
where $\bar{\theta}_{l,m}$ and $\bar{\phi}_{l,m}\in[-1,1]$ denote the spatial directions at the BS side and the user side of the $l$-th path component at the $m$-th subcarrier, respectively.
The spatial directions are the directions of the channel path components in the spatial domain. Since (3) is obtained by the DFT of (2), we can obtain the relationship between the spatial directions ($\bar{\theta}_{l,m},\bar{\phi}_{l,m}$) in (3) and the physical directions ($\tilde{\theta}_{l},\tilde{\phi}_{l}$) in (2) as $\bar{\theta}_{l,m}=2d\frac{f_{m}}{c}\sin{\tilde{\theta}_{l}} $ and $\bar{\phi}_{l,m}=2d\frac{f_{m}}{c}\sin{\tilde{\phi}_{l}}$. For simplification, in this paper, we use $\theta_{l}=\sin{\tilde{\theta}_{l}}$ and $\phi_{l}=\sin{\tilde{\phi}_{l}}$ to denote the physical directions, where $\theta_{l},\phi_{l}\in[-1,1]$.

\section{Delay-Phase Precoding for THz Massive MIMO}\label{Squint}

In this section, we will first introduce the beamforming mechanism in massive MIMO systems. Then, the beam split effect in THz massive MIMO will be revealed. After that, the DPP architecture will be proposed to mitigate the performance loss caused by the beam split effect.

\begin{figure}
	\centering
	\includegraphics[width=0.45\textwidth]{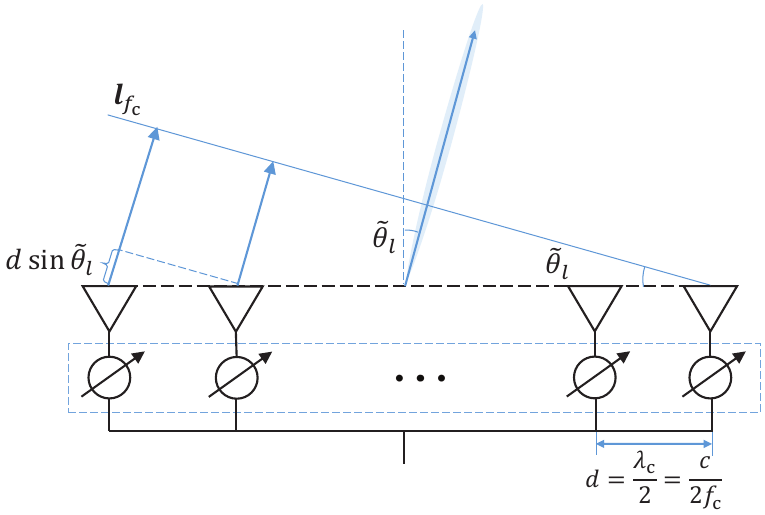}
	\caption{Beamforming mechanism.}
\end{figure}
\vspace{-3mm}
\subsection{Beamforming mechanism}\label{beam}
Generally, in THz massive MIMO systems, the analog beamformer is designed to generate beams towards the physical directions of channel path components to compensate for the severe path loss \cite{Ref:SpatiallyPre2014}. Then, the digital precoder is designed based on the determined analog beamformer to realize spatial multiplexing gain. Hence, whether the beams can precisely point to the physical directions of channel path components has a crucial impact on the achievable rate performance. Taking the narrowband system as an example, we will introduce the beamforming mechanism.

We consider the $l$-th path component with the physical direction $\theta_{l}$ of the channel in (\ref{3}). Usually, the $l$-th column of the analog beamformer $\mathbf{A}$ in (\ref{1}), i.e., the analog beamforming vector $\mathbf{a}_{l}=\mathbf{A}_{[:,l]}$, is used to generate a directional beam towards the $l$-th path's physical direction $\theta_{l}$, which has been proved to be near-optimal\cite{Ref:SpatiallyPre2014}. Specifically, the beamforming mechanism is to make electromagnetic waves that transmitted by different antenna elements form an equiphase surface, which
is perpendicular to the target physical direction $\theta_{l}$, as shown in Fig. 2. To achieve this goal, different phase shifts provided by PSs should be compensated at different antenna elements. For instance, the distance difference between adjacent antenna elements reaching the equiphase surface is $d\sin{\tilde{\theta}_{l}}=d\theta_{l}$. Therefore, for narrowband systems where $f_{m}\approx f_\mathrm{c}$, the phase difference that should be compensated between adjacent antenna elements is $-2\pi \frac{d}{\lambda_\mathrm{c}}\theta_{l}=-2\pi \frac{d}{c}f_\mathrm{c}\theta_{l}$. As a result, the analog beamforming vector $\mathbf{a}_{l}$ is
\begin{equation}\label{6}
\begin{aligned}
\mathbf{a}_{l}&=\frac{1}{N_\mathrm{t}}\left[1,e^{-j2\pi\frac{d}{c}f_\mathrm{c}\theta_{l}},e^{-j2\pi 2\frac{d}{c}f_\mathrm{c}\theta_{l}},\cdots,e^{-j2\pi(N_\mathrm{t}-1)\frac{d}{c}f_\mathrm{c}\theta_{l}}\right]^{T}\\
&=\mathbf{f}_\mathrm{t}\left(2\frac{d}{c}f_\mathrm{c}\theta_{l}\right)=\mathbf{f}_\mathrm{t}\left(\theta_{l}\right).
\end{aligned}
\end{equation}

Based on the analog beamforming vector designed in (\ref{6}), the normalized array gain $\eta(\mathbf{a}_{l},\theta_{l},f_{c})$ achieved by $\mathbf{a}_{l}$ in the physical direction $\theta_{l}$ at the central frequency $f_\mathrm{c}$ is
\begin{equation}\label{7}
\begin{aligned}
\eta(\mathbf{a}_{l},\theta_{l},f_{c})&=\left|\mathbf{f}_\mathrm{t}(2d\frac{f_{c}}{c}\theta_{l})^{H}\mathbf{a}_{l}\right|\\
&\overset{(a)}{=}\left|\mathbf{f}_\mathrm{t}(\theta_{l})^{H}\mathbf{a}_{l}\right|\approx\left|\mathbf{f}_\mathrm{t}(\theta_{l})^{H}\mathbf{f}_\mathrm{t}(\theta_{l})\right|=1,
\end{aligned}
\end{equation}
where (a) comes from $d=\frac{\lambda_{c}}{2}=\frac{c}{2f_\mathrm{c}}$. It is clear from (\ref{7}) that by setting the analog beamforming vector $\mathbf{a}_{l}=\mathbf{f}_\mathrm{t}(\theta_{l})$, the optimal normalized array gain of $1$ can be achieved at the central frequency $f_\mathrm{c}$. Thus, considering $f_{m}\approx f_\mathrm{c},m=1,2,\cdots,M$, the narrowband systems can enjoy the satisfying normalized array gain across the entire bandwidth $B$. 
\vspace{-2mm}
\subsection{Beam split effect}\label{split}

In wideband systems, since the PSs are \emph{frequency-independent} and the antenna spacing $d$ is set according to the central frequency $f_\mathrm{c}$, the analog beamforming vector $\mathbf{a}_{l}$ is usually set the same as (\ref{6}), which is \emph{frequency-independent}. However, considering the frequency-independent phase shifts will cause frequency-dependent time delays, i.e., phase shift $\Delta\theta$ corresponds to time delay $-2\pi f_{m}\Delta\theta$ at subcarrier frequency $f_{m}$, the equiphase surfaces generated by the analog beamforming vector $\mathbf{a}_{l}$ will be separated at different subcarriers. Therefore, the beams generated by $\mathbf{a}_{l}$ will point to different physical directions surrounding the target physical direction $\theta_{l}$ at different subcarriers. This effect is called beam squint in mmWave massive MIMO systems \cite{Ref:WidebandEff2019}. Fortunately, since the beams can still cover the user by their mainlobes, the array gain degradation caused by the beam squint is small and can be solved by existing methods\cite{Ref:SubWideHy2017,Ref:WideMIMOHy2018,Ref:WideCode2016,Ref:WideCode2019}.

However, the beam squint effect will be severely aggregated in THz massive MIMO systems for two reasons. Firstly, due to the much larger bandwidth of THz massive MIMO systems, the deviations between the physical directions that the beams at different subcarriers are aligned with and the target physical direction will significantly increase. Secondly, the much larger number of antennas in THz massive MIMO systems leads to an extremely narrow beamwidth. Due to the above two reasons, the beams at different subcarrier frequencies may be totally split into separated physical directions in THz massive MIMO systems as shown in Fig. 3 (a). Thus, unlike the beam squint effect, an unacceptable array gain loss occurs since most of the beams at different subcarriers cannot cover the user in their mainlobes. Unfortunately, this aggregated effect cannot be solved by the existing methods based on the conventional hybrid precoding architecture\cite{Ref:SubWideHy2017,Ref:WideMIMOHy2018,Ref:WideCode2016,Ref:WideCode2019}, and it has not been well addressed in the literature for THz massive MIMO systems. To this end, we define the effect that the beams at different subcarriers are totally separated as the \textbf{beam split effect}. A simple analogy between the beam split effect and the rainbow can illustrate the mechanism of the beam split effect. As shown in Fig. 3 (b), because the tiny water droplets in the air have different refractive indices for the wideband white light, the pure light of different frequencies will totally separate and eventually produce the rainbow. Just like the rainbow, frequency-independent PSs cause different ``refractive indices" for THz signals at different frequencies, and thus leads to totally separated beams at different frequencies. Therefore, we can also call the beam split effect as ``\textbf{THz rainbow}".

The following \textbf{Lemma 1} will theoretically quantify the severe array gain loss caused by the beam split effect, which is mainly determined by the bandwidth and the number of antennas.

\begin{figure*}
	\centering
	\includegraphics[width=0.97\textwidth]{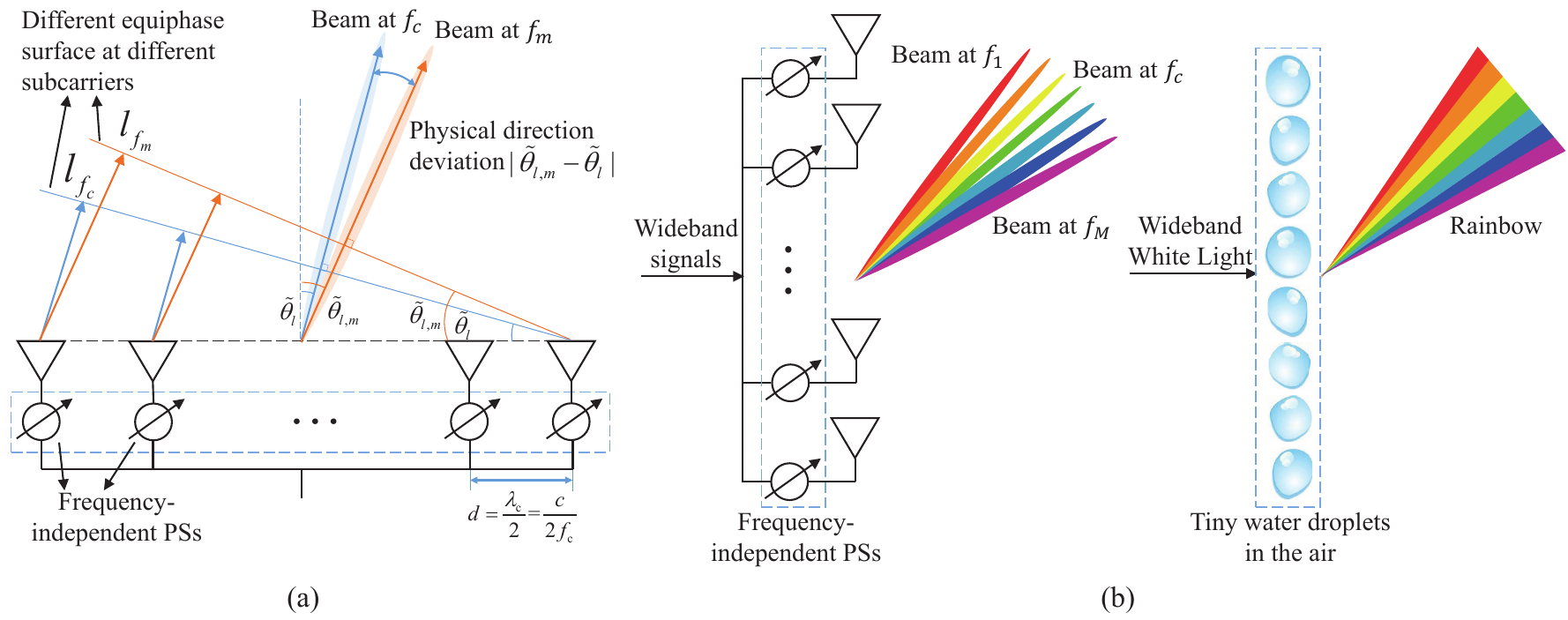}
	\caption{The mechanism of the beam split effect: (a) The beam split at subcarrier frequency $f_{m}$; (b) The analogy between the beam split effect and the rainbow over the large bandwidth. }
\end{figure*}

\newtheorem{thm}{Lemma}
\begin{thm}
The beam generated by the analog beamforming vector $\mathbf{a}_{l}=\mathbf{f}_\mathrm{t}(\theta_{l})$ is aligned with the physical direction $\theta_{l,m}$ satisfying $\theta_{l,m}=\frac{\theta_{l}}{\xi_{m}}$, where $\xi_{m}=\frac{f_{m}}{f_\mathrm{c}}$ is the relative frequency. Furthermore, when $\left|(\xi_{m}-1)\theta_{l}\right|\geq\frac{2}{N_\mathrm{t}}$, the normalized array gain $\eta(\mathbf{a}_{l},\theta_{l},f_{m})$ achieved by the analog beamforming vector $\mathbf{a}_{l}$ in the physical direction $\theta_{l}$ at subcarrier frequency $f_{m}$ satisfies
\begin{equation}
\eta(\mathbf{a}_{l},\theta_{l},f_{m})\leq\frac{1}{N_\mathrm{t}\sin{\frac{3\pi}{2N_\mathrm{t}}}}.
\end{equation}
\end{thm}
\emph{Proof:} The normalized array gain achieved by the analog beamforming vector $\mathbf{a}_{l}$ in an arbitrary physical direction $\theta\in[-1,1]$ at subcarrier frequency $f_{m}$ can be denoted as
$\eta(\mathbf{a}_{l},\theta,f_{m})=\left|\mathbf{f}_\mathrm{t}(2d\frac{f_{m}}{c}\theta)^{H}\mathbf{a}_{l}\right|$.
Then, we have
\begin{equation}\label{10}
\begin{aligned}
\eta(\mathbf{a}_{l},\theta,f_{m})&\overset{(a)}=\left|\mathbf{f}_\mathrm{t}(2d\frac{f_{m}}{c}\theta)^{H}\mathbf{f}_\mathrm{t}(\theta_{l})\right|\\
&\overset{(b)}{=}\frac{1}{N_\mathrm{t}}\left|\sum_{n=0}^{N_{t}-1}e^{jn\pi(\xi_{m}\theta-\theta_{l})}\right|\\
&\overset{(c)}{=}\left|\frac{\sin{\frac{N_{t}\pi}{2}}(\xi_{m}\theta-\theta_{l})}{N_{t}\sin{\frac{\pi}{2}}(\xi_{m}\theta-\theta_{l})}e^{-j\frac{(N_{t}-1)\pi(\xi_{m}\theta-\theta_{l})}{2}}\right|,
\end{aligned}
\end{equation}
where (a), (b), and (c) come from $\mathbf{a}_{l}=\mathbf{f}_\mathrm{t}(\theta_{l})$, (\ref{4}), and the equation $\Sigma_{n=0}^{N-1}e^{jn\pi\alpha}=\frac{\sin{\frac{N_{t}\pi}{2}}\alpha}{N_\mathrm{t}\sin{\frac{\pi}{2}}\alpha}e^{-j\frac{(N-1)\pi}{2}\alpha}$, respectively. Then, we can obtain
\begin{equation}\label{11}
\eta(\mathbf{a}_{l},\theta,f_{m})=\frac{1}{N_\mathrm{t}}|\Xi_{N_{t}}((\xi_{m}\theta-\theta_{l}))|,
\end{equation}
where $\Xi_{N_\mathrm{t}}(x)=\left(\sin{\frac{N_{t}\pi}{2}x}\right)/\left(\sin{\frac{\pi}{2}}x\right)$ is the Dirichlet sinc function \cite{Ref:DeconMIMO2002}. It is known that the Dirichlet sinc function has the power-focusing property, where the maximum is $|\Xi_{N_\mathrm{t}}(0)|=N_{t}$ and the value of $|\Xi_{N_\mathrm{t}}(x)|$ decreases sharply as $|x|$ increases\cite{Ref:DeconMIMO2002}. 

Denote $\theta_{l,m}$ as the physical direction that the analog beamforming vector $\mathbf{a}_{l}$ is aligned with at subcarrier frequency $f_{m}$. The analog beamforming vector $\mathbf{a}_{l}$ should achieve the largest array gain in the physical direction $\theta_{l,m}$ when $\theta_{l,m}=\arg\max_{\theta}\eta(\mathbf{a}_{l},\theta,f_{m})$. Therefore, considering the maximum of $|\Xi_{N_\mathrm{t}}(x)|$ is $|\Xi_{N_\mathrm{t}}(0)|=N_{t}$, the physical direction $\theta_{l,m}$ should satisfy $\xi_{m}\theta_{l,m}-\theta_{l}=0$ according to (\ref{11}). Hence, we can obtain
\begin{equation}\label{12}
\theta_{l,m}=\theta_{l}/\xi_{m}.
\end{equation}
Moreover, by substituting $\theta=\theta_{l}$ into (\ref{11}), the normalized array gain $\eta(\mathbf{a}_{l},\theta_{l},f_{m})$ can be denoted as $\eta(\mathbf{a}_{l},\theta_{l},f_{m})=\frac{1}{N_\mathrm{t}}|\Xi_{N_{t}}((\xi_{m}-1)\theta_{l})|$. According to the power-focusing property of the Dirichlet sinc function, when $|(\xi_{m}-1)\theta_{l}|\geq\frac{2}{N_\mathrm{t}}$, $|(\xi_{m}-1)\theta_{l}|$ locates out of the mainlobe of the Dirichlet sinc function $|\Xi_{N_{t}}(x)|$. This indicates the beam generated by the analog beamforming  vector $\mathbf{a}_{l}$ cannot cover the user with its mainlobe, and a severe array gain loss will occur. Considering that the maximum value of $|\Xi_{N_\mathrm{t}}(x)|$ when $|x|\geq\frac{2}{N_\mathrm{t}}$ is $(1/N_\mathrm{t})\sin{\frac{3\pi}{2N_\mathrm{t}}}$, we have 
\begin{equation}\label{13}
\left|\eta(\mathbf{a}_{l},\theta_{l},f_{m})\right|=\frac{1}{N_{t}}|\Xi_{N_\mathrm{t}}((\xi_{m}-1)\theta_{l})|\leq\frac{1}{N_\mathrm{t}\sin{\frac{3\pi}{2N_\mathrm{t}}}},
\end{equation}
when $|(\xi_{m}-1)\theta_{l}|\geq\frac{2}{N_\mathrm{t}}$.$\hfill\blacksquare$

\textbf{Lemma 1} has revealed two major parameters that affect the array gain loss caused by the beam split effect, i.e., the bandwidth $B$ and the number of antennas $N_\mathrm{t}$. Specifically, when the bandwidth $B$ and the number of antennas $N_\mathrm{t}$ are large, the condition $|(\xi_{m}-1)\theta_{l}|\geq\frac{2}{N_\mathrm{t}}$ can be easily satisfied at most subcarriers, which indicates that beams at most of subcarriers  cannot cover the user with their mainlobes. As a result, the achieved array gains at most of the subcarriers are upper bounded by $\frac{1}{N_\mathrm{t}\sin{\frac{3\pi}{2N_\mathrm{t}}}}$ as shown in (\ref{13}), which are extremely small. This array gain loss caused by the beam split effect can be explained from the physical perspective. Firstly, when the bandwidth $B$ grows up, the physical direction deviation, i.e., $|\theta_{l,m}-\theta_{l}|$, increases. Secondly, a larger number of antennas $N_\mathrm{t}$ will result in a narrower beamwidth. Consequently, considering the large bandwidth $B$ and the large antenna number $N_\mathrm{t}$ in THz massive MIMO systems, there will be a large physical direction deviation and an extremely narrow beamwidth at most of the subcarrier frequencies. Under this case, the beams at different subcarriers will become totally separated, which means the beam split effect occurs and thus cause a severe array gain loss. 

\begin{figure*}
	\centering
	\includegraphics[width=0.97\textwidth]{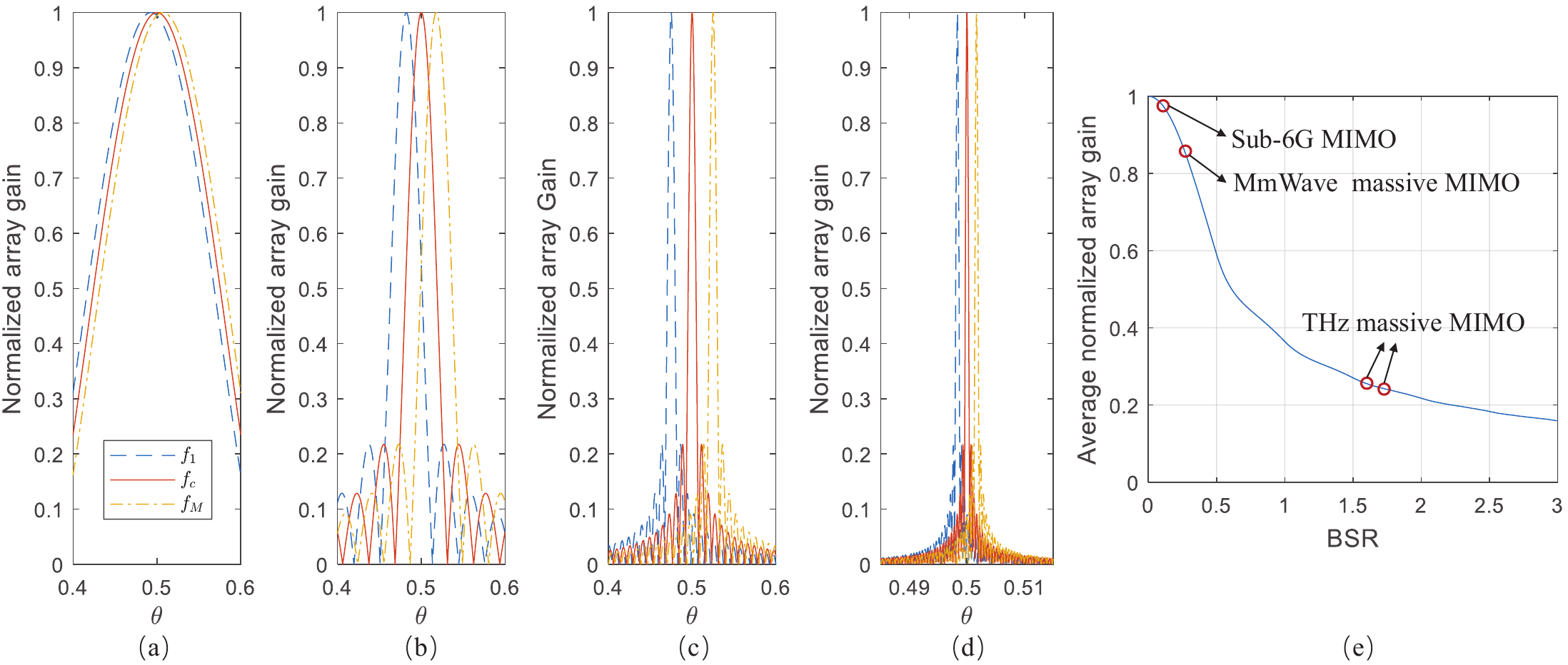}
	\caption{Comparison of normalized array gain achieved by the conventional hybrid precoding architecture with analog beamforming vector $\mathbf{a}_{l}=\mathbf{f}_{t}(\theta_{l})$, $\theta_{l}=0.5$ and $M=128$: (a) Sub-6G MIMO with $f_\mathrm{c}=3.5$ GHz, $B=0.1$ GHz, and $N_{t}=16$; (b) MmWave massive MIMO with $f_\mathrm{c}=28$ GHz, $B=2$ GHz, and $N_{t}=64$; (c) THz massive MIMO with $f_\mathrm{c}=300$ GHz, $B=30$ GHz, and $N_{t}=256$; (d) THz massive MIMO with $f_\mathrm{c}=300$ GHz, $B=2$ GHz, and $N_{t}=4096$; (e) Average nomalized array gain with respect to the BSR defined in (\ref{13-3}), where we mark points corresponding to the sub-6G MIMO in (a), mmWave massive MIMO in (b), and THz massive MIMO in (c) and (d), respectively. }
\end{figure*}

\begin{figure}
	\centering
	\includegraphics[width=0.47\textwidth]{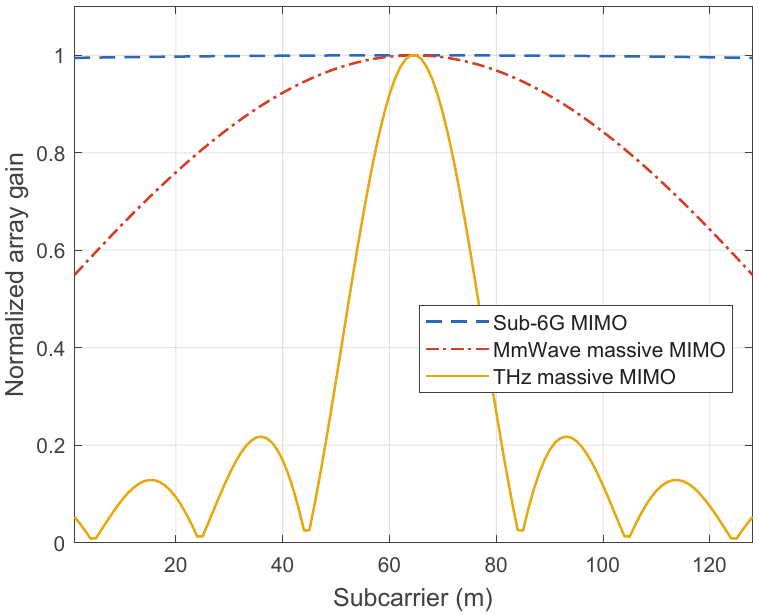}
	\caption{Normalized array gain achieved by the conventional hybrid precoding architecture with analog beamforming vector $\mathbf{a}_{l}=\mathbf{f}_{t}(\theta_{l})$ in different systems. The system parameters are the same as these in Fig. 4 (a), (b) and (c).}
\end{figure}

It is important to define a simple metric to evaluate the degree of the beam split effect. Actually, from the description above, we can conclude that the degree of the beam split effect is determined by the ``relative offset" between the physical direction deviation and the beamwidth. When the relative offset is large, beams tend to split at different subcarriers. While, when the relative offset is small, beams will not even squint. Following this idea, we define a metric called beam split ratio (BSR). Specifically, the BSR is defined as the expectation of the ratio between the physical direction deviation and half of the beamwidths for all subcarrier frequencies and physical directions as
\begin{equation}\label{13-3}
\begin{aligned}
\mathrm{BSR}&=\frac{1}{2M}\int_{-1}^{1}\sum_{m=1}^{M}\frac{|\theta_{l,m}-\theta_{l}|}{2/N_\mathrm{t}}\mathrm{d}\theta_{l}\\
&=\int_{-1}^{1}\frac{1}{M}\sum_{m=1}^{M}\frac{1}{2}|N_\mathrm{t}(\xi_{m}-1)\theta_{l}|\mathrm{d}\theta_{l},
\end{aligned}
\end{equation}
where $|\theta_{l,m}-\theta_{l}|$ is the physical direction deviation, and $2/N_\mathrm{t}$ denotes half of the beamwidth. Note that the condition $|(\xi_{m}-1)\theta_{l}|\geq\frac{2}{N_\mathrm{t}}$ or $\frac{1}{2}|N_\mathrm{t}(\xi_{m}-1)\theta_{l}|\geq 1$ means the beam at the subcarrier frequency $f_{m}$ cannot cover the user with its mainlobe, and the BSR is defined as the average of $\frac{1}{2}|N_\mathrm{t}(\xi_{m}-1)\theta_{l}|$. Therefore, we can suppose that if $\mathrm{BSR}> 1$, the beams at different subcarriers cannot cover the user with their mainlobes on average and the beam split effect occurs. On the contrary, when $\mathrm{BSR}<1$, the beams will only slightly squint. A larger BSR means a stronger degree of the beam split effect and a more serious array gain loss. For instance, for a THz massive MIMO system with parameters $f_\mathrm{c}=300$ GHz, $N_\mathrm{t}=256$, $M=128$ and $B=30$ GHz, $\mathrm{BSR}=1.6> 1$\footnote{We consider the THz massive MIMO system with a bandwidth of $30$ GHz to support Tbps ultra-high-speed applications\cite{Ref:To6G2020}. For example, a data rate of about $0.58$ Tbps is required for mobile hologram\cite{Ref:Holo2011}. Such a data rate requires at least $36.25$ GHz bandwidth with 256 quadrature amplitude modulation (QAM) or $32.22$ GHz bandwidth with 512 QAM, so it is reasonable to assume a $30$ GHz bandwidth for THz massive MIMO.}, which indicates the beam split effect happens in THz massive MIMO systems. While, for a mmWave massive MIMO, the BSR is usually smaller than $1$, e.g., $\mathrm{BSR}=0.29$ for a mmWave massive MIMO system with parameters $f_\mathrm{c}=28$ GHz, $N_\mathrm{t}=64$, $M=128$ and $B=2$ GHz. This indicates that the beam squint rather than the beam split happens in mmWave massive MIMO. Considering the performance loss caused by the beam squint is limited compared with that caused by the beam split, we can conclude that the beam split effect is one of the differences between mmWave and THz massive MIMO systems.

To better illustrate the beam split effect, we compare the normalized array gain comparison achieved by the analog beamforming vector $\mathbf{a}_{l}$ between sub-6G MIMO system, mmWave massive MIMO system, and THz massive MIMO system in Fig. 4 (a), (b), (c), and (d). The BSRs of them are $0.03$, $0.29$, $1.6$, and $1.71$, respectively. In Fig. 4 (e), we provide the average normalized array gain with respect to the defined BSR. We can observe from Fig. 4 (a) and (b) that in sub-6G and mmWave systems with $\mathrm{BSR}<1$, the beams at subcarrier frequencies $f_{1}$ and $f_{M}$ slightly squint from the beam at the central frequency $f_\mathrm{c}$. By contrast, for THz massive MIMO systems in Fig. 4 (c) and (d) with $\mathrm{BSR}>1$, the beams at subcarrier frequencies $f_{1}$ and $f_{M}$ are totally separated from the beam at the central frequency $f_\mathrm{c}$. Fig. 4 (c) illustrates that a wide bandwidth $B$ will cause severe beam split effect. Fig. 4 (d) shows that even if the bandwidth is not large, i.e., equal to that in mmWave systems in Fig. 4 (b), a large antenna number $N_\mathrm{t}$ will also introduce the serious beam split. The above observations are consistent with the array gain analysis and the defined BSR highlighted in Fig. 4 (e), which verifies that the beam split effect is caused by the wide bandwidth and large antenna number in THz massive MIMO systems. Moreover, Fig. 4 (e) shows that the average normalized array gain decreases monotonically with BSR, which indicates the defined BSR could be able to efficiently reflect the degree of the beam split effect. Besides, Fig. 5 illustrates the normalized array gain achieved by the analog beamforming vector $\mathbf{a}_{l}$ at different subcarriers $m$. We can see from Fig. 5 that the analog beamforming vector $\mathbf{a}_{l}$ suffers from a severe array gain loss in wideband THz massive MIMO system due to the beam split effect. While, the array gain losses in sub-6G and mmWave systems are not serious. Particularly, for more than $50\%$ of subcarriers, e.g., the subcarriers $m\leq47$ or $m\geq81$ in THz massive MIMO system, the user will suffer from more than $80\%$ array gain loss. 

Such a serious array gain loss incurred by the beam split effect is not acceptable for THz communications. However, the existing hybrid precoding methods with frequency-independent PSs cannot solve this problem. Several TTD based architectures have been proposed to solve the beam split\cite{Ref:TTD2008,Ref:TTD2019,Ref:TTDTrain2020,Ref:HybridTTD2021}, where each antenna element is controlled by one TTD. But these solutions are unpractical for massive MIMO systems, since a large number of TTDs may bring high power consumption and high hardware cost\cite{Ref:WidebandEff2019}. To our best knowledge, there is no practical precoding architecture and corresponding precoding design to solve the beam split effect in THz massive MIMO systems. To fill in this gap, in the next subsection we will propose a new precoding architecture called DPP for THz massive MIMO systems.

\subsection{Delay-phase precoding (DPP)}\label{subBeam}
As discussed in Subsection \ref{split} above, due to the beam split effect, the \emph{frequency-independent} beamformer generated by the frequency-independent PSs in the conventional hybrid precoding architecture, will result in severe array gain loss. In this subsection, we will propose a new precoding architecture called DPP to solve this problem. As shown in Fig. 6, compared with the conventional hybrid precoding architecture, a TD network is introduced as a new precoding layer between the RF chains and the frequency-independent PS network in the proposed DPP. Specifically, each RF chain is connected to $K$ TD elements, and then each TD element is connected to $P=\frac{N_\mathrm{t}}{K}$ PSs in a sub-connected manner\cite{Ref:EnHP2016}. Therefore, each RF chain still connects to every antenna element through the PSs. The TD network can realize frequency-dependent phase shifts through time delays, e.g., the phase shift $-2\pi f_{m}t$ can be achieved by the time delay $t$ at the subcarrier frequency $f_{m}$. Thus, by utilizing the TD network, the proposed DPP converts the traditional phase-controlled beamformer into delay-phase jointly controlled beamformer, which can realize the \emph{frequency-dependent} beamforming. 
\begin{figure*}
	\centering
	\includegraphics[width=0.9\textwidth]{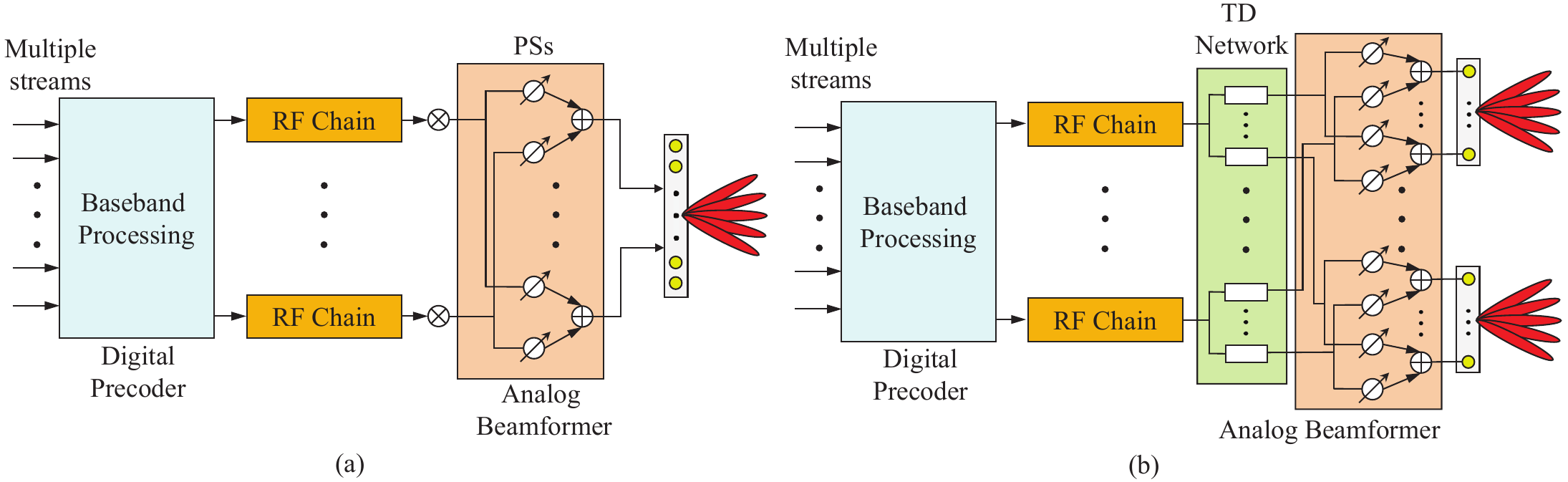}
	\caption{Precoding architecture comparison:  (a) The conventional hybrid precoding architecture for mmWave massive MIMO; (b) The proposed DPP architecture for THz massive MIMO.}
\end{figure*}

Without loss of generality, we consider the $l$-th channel path component. Since the TD network can provide the frequency-dependent phase shifts, we now utilize the frequency-dependent $\mathbf{a}_{l,m}$ instead of the frequency-independent $\mathbf{a}_{l}$ to represent the analog beamforming vector generated by the DPP for the $l$-th path component at the $m$-th subcarrier. $\mathbf{a}_{l,m}$ can be denoted as
\begin{equation}\label{13-1}
\mathbf{a}_{l,m}=
\mathrm{blkdiag}\big(\left[\bar{\mathbf{a}}_{l,1}, 
\bar{\mathbf{a}}_{l,2},\cdots, \bar{\mathbf{a}}_{l,K}\right]\big) 
\mathbf{p}_{l,m},
\end{equation}
where $\bar{\mathbf{a}}_{l,k}\in\mathcal{C}^{P\times1}$ with $k=1,2,\cdots,K$ denotes the analog beamforming vector realized by PSs connected to the $k$-th TD element, so we have $|\bar{\mathbf{a}}_{l,k,[j]}|=\frac{1}{\sqrt{N_{t}}}$ as usual due to the constraint of constant modulus, and $\mathbf{p}_{l,m}\in\mathcal{C}^{K\times1}$ composes of the frequency-dependent phase shift realized by $K$ TD elements. Specifically, the $k$-th element $\mathbf{p}_{l,m,[k]}$ in $\mathbf{p}_{l,m}$ with $k=1,2,\cdots,K$ satisfies the form $\mathbf{p}_{l,m,[k]}=e^{-j2\pi f_{m}t_{l,k}}$, where the time delay provided by the $k$-th TD element is $t_{l,k}$. 

Aiming to compensate for the severe array gain loss caused by the beam split effect, the analog beamforming vector $\mathbf{a}_{l,m}$ should generate beams aligned with the target physical direction $\theta_{l}$ at all $M$ subcarriers. To realize this goal, we will first use the frequency-independent PSs to generate a beam aligned with the target physical direction $\theta_{l}$ as
\begin{equation}\label{14}
[\bar{\mathbf{a}}_{l,1}^{T},\bar{\mathbf{a}}_{l,2}^{T},\cdots,\bar{\mathbf{a}}_{l,K}^{T}]^{T}=\mathbf{f}_\mathrm{t}(\theta_{l}).
\end{equation}
Then, we utilize the frequency-dependent $\mathbf{p}_{l,m}$ in (\ref{13-1}) to rotate the physical direction that the beam  $[\bar{\mathbf{a}}_{l,1}^{T},\bar{\mathbf{a}}_{l,2}^{T},\cdots,\bar{\mathbf{a}}_{l,K}^{T}]^{T}=\mathbf{f}_\mathrm{t}(\theta_{l})$ is aligned with from $\theta_{l,m}$ to $\theta_{l}$. To maintain the directivity of the frequency-dependent beam generated by the analog beamforming vector $\mathbf{a}_{l,m}$, we set $\mathbf{p}_{l,m}$ share the same form as the array response in (\ref{4}). Specifically, by making the frequency-dependent phase shift $-2 f_{m}t_{l,k}=-(k-1)\beta_{l,m}$ with $k=1,2,\cdots,K$, $\mathbf{p}_{l,m}$ satisfies
\begin{equation}\label{15}
\mathbf{p}_{l,m}=[1,e^{-j\pi\beta_{l,m}},e^{-j2\pi\beta_{l,m}},\cdots,e^{-j\pi(K-1)\beta_{l,m}}]^{T},
\end{equation}
where we define $\beta_{l,m}$ as the direction rotation factor at the $m$-th subcarrier for the $l$-th path component. We set the value range of the direction rotation factor as $\beta_{l,m}\in[-1,1]$, due to the periodicity of the $\mathbf{p}_{l,m}$ in (\ref{15}) with respect to the direction rotation factor $\beta_{l,m}$. By adjusting the direction rotation factor $\beta_{l,m}$, the beam generated by the analog beamforming vector $\mathbf{a}_{l,m}$ can be made to be aligned with the target physical direction $\theta_{l}$ at all $M$ subcarriers, which is proved by the following \textbf{Lemma 2}.

\begin{thm}
	When $\bar{\mathbf{a}}_{l,k}$ satisfies $[\bar{\mathbf{a}}_{l,1}^{T},\bar{\mathbf{a}}_{l,2}^{T},\cdots,\bar{\mathbf{a}}_{l,K}^{T}]^{T}=\mathbf{f}_\mathrm{t}(\theta_{l})$ as shown in (\ref{14}) and $\mathbf{p}_{l,m}=[1,e^{j\pi\beta_{l,m}},\cdots,e^{j\pi(K-1)\beta_{l,m}}]^{T}$ as shown in (\ref{15}), the beam generated by the analog beamforming vector $\mathbf{a}_{l,m}$ can be aligned with the physical direction $\theta_\mathrm{opt}$ at the subcarrier frequency $f_{m}$ as
	\begin{equation}\label{16}
	\theta_\mathrm{opt}=\arg\max_{\theta}|\eta({\mathbf{a}}_{l,m},\theta,f_{m})|=\frac{\theta_{l}}{\xi_{m}}+\frac{\beta_{l,m}}{\xi_{m}P},
	\end{equation}
	where $P=N_\mathrm{t}/K$, and the normalized array gain achieved by the analog beamforming vector $\mathbf{a}_{l,m}$ in the physical direction $\theta_\mathrm{opt}$ is $\eta({\mathbf{a}}_{l,m},\theta_\mathrm{opt},f_{m})=\frac{K}{N_\mathrm{t}}\Xi_{P}(\frac{\beta_{l,m}}{P})$.
\end{thm}

\emph{Proof}: See Appendix.

We can know from \textbf{Lemma 2} that the direction rotation factor $\beta_{l,m}$ can change the physical direction of the beam from $\theta_{l,m}=\frac{\theta}{\xi_{m}}$ achieved by the conventional hybrid precoding architecture to $\theta_\mathrm{opt}=\frac{\theta_{l}}{\xi_{m}}+\frac{\beta_{l,m}}{\xi_{m}P}$. Therefore, to compensate for the array gain loss caused by the beam split effect across the entire bandwidth, we should make $\theta_\mathrm{opt}=\theta_{l}$, i.e.,
\begin{equation}\label{22}
\frac{\theta_{l}}{\xi_{m}}+\frac{\beta_{l,m}}{\xi_{m}P}=\theta_{l}.
\end{equation}
Then, we can easily obtain the direction rotation factor $\beta_{l,m}$ as
\begin{equation}\label{23}
\beta_{l,m}=(\xi_{m}-1)P\theta_{l}.
\end{equation} 
By setting the direction rotation factor $\beta_{l,m}$ as (\ref{23}), the array gain loss incurred by the beam split effect can be efficiently eliminated, since the beams generated by the analog beamforming vector $\mathbf{a}_{l,m}$ are aligned with the physical direction $\theta_{l}$ across the entire bandwidth. As we can see from the Appendix, the beamwidths of beams generated by $\mathbf{a}_{l,m}$ are approximately decided by $|\Xi_{K}(P(\theta_{l}-\xi_{m}\theta)+\beta_{l,m})|$, which are equal to the original beam generated by $\mathbf{a}_{l}=\mathbf{f}_\mathrm{t}(\theta_{l})$ as $\frac{4}{N_{t}}$.

An important problem in the proposed DPP is that how many TD elements are sufficient to mitigate the beam split at all subcarriers in all possible physical directions. Note that the value of the direction rotation factor is restricted by $\beta_{l,m}\in[-1,1]$. Thus, the direction rotation factor $\beta_{l,m}$ calculated as (\ref{23}) should lie in $[-1,1]$ in all possible physical directions at all subcarrier frequencies $f_{m}$. This means $-1\leq(\xi_{m}-1)P\theta_{l}\leq1$. Recalling that $\theta_{l}\in[-1,1]$ and $\frac{f_{1}}{f_\mathrm{c}}\leq\xi_{m}\leq\frac{f_{M}}{f_\mathrm{c}}$, we have $P\leq\frac{1}{f_{M}/{f_\mathrm{c}}-1}$. Substituting $K=N_\mathrm{t}/P$ into $P\leq\frac{1}{f_{M}/{f_\mathrm{c}}-1}$, we can obtain the constraint on the number of the TD elements $K$ as
\begin{equation}\label{26}
K\geq(f_{M}/{f_\mathrm{c}}-1)N_\mathrm{t}.
\end{equation}
From (\ref{26}), we can observe that the number of TD elements $K$ increases linearly with the ratio between the maximum subcarrier frequency $f_{M}$ and the central frequency $f_\mathrm{c}$. Since $f_{M}/f_\mathrm{c}$ is proportional to the bandwidth $B$, we can know that $K$ increases linearly with the bandwidth $B$.  On the other hand, in the narrowband system with the assumption $f_{m}\approx f_\mathrm{c}$, $K$ becomes $0$. This means that the proposed DPP architecture degenerates into the conventional hybrid precoding architecture in the narrowband case. In practical systems, since $f_{M}/f_\mathrm{c}$ usually has an upper bound due to hardware constraint, e.g., power amplifier\cite{Ref:PALimite2019}, we can design the number of TD elements $K$ based on the upper bound of $f_{M}/f_\mathrm{c}$. Thus, when the central frequency $f_\mathrm{c}$ is fixed, (\ref{26}) can be always satisfied with smaller $f_{M}$, i.e., smaller bandwidths. In this way, the proposed DPP architecture could adapt to systems with different bandwidths by using an identical number of TD elements $K$. Meanwhile, we should emphasize that even if we set the number of TD elements $K$ according to the upper bound of $f_{M}/{f_\mathrm{c}}$, the number of TD elements $K$ is still much smaller than the antenna number $N_\mathrm{t}$. For example, when $f_{M}/{f_\mathrm{c}}\leq 1.05$ and $N_\mathrm{t}=256$, $K$ should satisfy $K>12.8$. Then, considering that $P=\frac{N_\mathrm{t}}{K}$ should be an integer, we can set $K=16$ which is much smaller than $N_\mathrm{t}=256$. Notice that such a small number of TD elements introduce relatively small power consumption, and make the proposed DPP architecture be able to adapt to a wide frequency range.

\vspace{-3mm}
\subsection{Array gain performance of DPP}\label{array}
In this subsection, we will provide the theoretical analysis of the array gain achieved by the proposed DPP architecture. From (\ref{21}), for the $m$-th subcarrier, we have
\begin{equation}\label{27}
\eta(\mathbf{a}_{l,m},\theta_{l},f_{m})=\frac{K}{N_\mathrm{t}}\left|\Xi_{P}(\frac{\beta_{l,m}}{P})\right|,
\end{equation}
where the $\theta_\mathrm{opt}$ is replaced by $\theta_{l}$, because we have $\theta_\mathrm{opt}=\theta_{l}$ due to the proposed DPP. By substituting $\beta_{l,m}=(\xi_{m}-1)P\theta_{l}$ in (\ref{23}) into (\ref{27}), the expectation of the array gain achieved by $\mathbf{a}_{l,m}$ at all subcarriers in all possible physical direction can be denoted as
\begin{equation}\label{28}
\mathbb{E}(\eta(\mathbf{a}_{l,m},\theta_{l},f_{m})=\frac{K}{2MN_\mathrm{t}}\sum_{m=1}^{M}\int_{-1}^{1}|\Xi_{P}((\xi_{m}-1)\theta_{l})|\mathrm{d}\theta_{l}.
\end{equation}
As it is difficult to calculate the integration of the Dirichlet sinc function, we utilize a polynomial to fit it by three points $(-1,|\Xi_{P}(1-\xi_{m})|)$, $(0,P)$ and $(1,|\Xi_{P}(\xi_{m}-1)|)$, which leads
\begin{equation}\label{29}
\begin{aligned}
\int_{-1}^{1}&{|\Xi_{P}((\xi_{m}-1)\theta_{l})|\mathrm{d}\theta_{l}}\\
&\approx\int_{-1}^{1}\left[(\Xi_{P}(\xi_{m}-1)-P)\theta_{l}^{2}+P\right]\mathrm{d}\theta_{l}\\
&=\frac{2}{3}|\Xi_{P}(\xi_{m}-1)|+\frac{4}{3}P.
\end{aligned}
\end{equation}
By substituting (\ref{29}) into (\ref{28}), we have
\begin{equation}\label{30}
\mathbb{E}(\eta(\mathbf{a}_{l,m},\theta_{l},f_{m}))\approx\frac{K}{MN_\mathrm{t}}\sum_{m=1}^{M}\left(\frac{1}{3}|\Xi_{P}(\xi_{m}-1)|+\frac{2}{3}P\right).
\end{equation}

It is clear from (\ref{30}) that the expectation of the array gain achieved by the proposed DPP is mainly decided by the relative frequency $\xi_{m}$. Considering the constraint of (\ref{26}), $\xi_{m}-1$ always locates in the mainlobe of the Dirichlet sinc function $\Xi_{P}$. This guarantees the array gain achieved by the DPP is larger than $\frac{2KP}{3N_{t}}=0.667$, which is much higher than the array gain achieved by the conventional hybrid precoding architecture as shown in Fig. 5. For instance, when $f_\mathrm{c}=300$ GHz, $B=15$ GHz, $M=128$, $K=8$ and $N_{t}=256$, we have $\mathbb{E}(|\eta(\mathbf{a}_{l,m},\theta_{l},f_{m})|)\approx0.96$, which means the proposed DPP is able to approach the near-optimal array gain.

\section{Hardware Implementation of the DPP}\label{SPOB}

The hardware implementation of the DPP architecture is important to make the DPP concept practical in real THz massive MIMO systems. In this section, we will propose a practical hardware structure to realize the concept of the DPP based on TTDs. 

\subsection{True-time-delayers based DPP}\label{TTD}
Based on the intuitive idea that utilizing TTDs can be directly used to realize the TD network in the proposed  DPP architecture, we propose a hardware structure called TTD-DPP, as shown in Fig. 7. In the TTD-DPP structure, each RF chain is connected to $K$ TTDs, and each TTD is connected to $P=\frac{{N}_\mathrm{t}}{K}$ PSs. The TTDs can realize the phase shift $-2\pi f_{m}t$ by the time delay $t$ at frequency $f_{m}$. Therefore, the received signal $\mathbf{y}_{m}$ at the $m$-th subcarrier in (\ref{1}) can be denoted as
\begin{equation}\label{31}
\mathbf{y}_{m}=\sqrt{\rho}\mathbf{H}_{m}^{H}\mathbf{A}_\mathrm{u}\mathbf{A}_{m}^\mathrm{TTD}\mathbf{D}_{m}\mathbf{s}_{m}+\mathbf{n}_{m},
\end{equation}
where $\mathbf{A}_\mathrm{u}\in\mathcal{C}^{N_\mathrm{t}\times KN_\mathrm{RF}}$ denotes the analog beamformer provided by the frequency-independent PSs with $\mathbf{A}_\mathrm{u}=[\mathbf{A}_{\mathrm{u},1},\mathbf{A}_{\mathrm{u},2},\cdots,\mathbf{A}_{\mathrm{u},N_\mathrm{RF}}]$. $\mathbf{A}_{\mathrm{u},l}=\mathrm{blkdiag}([\bar{\mathbf{a}}_{l,1},\bar{\mathbf{a}}_{l,1},\cdots,\bar{\mathbf{a}}_{l,K}])$ denotes the analog beamformer realized by the PSs connected to the $l$-th RF chain through TTDs. $\mathbf{A}_{m}^\mathrm{TTD}\in\mathcal{C}^{KN_\mathrm{RF}\times N_\mathrm{RF}}$ denotes the frequency-dependent phase shifts realized by TTDs, which satisfies $\mathbf{A}_{m}^\mathrm{TTD}=\mathrm{blkdiag}\left( 
[e^{-j2\pi f_{m}\mathbf{t}_{1}},e^{-j2\pi f_{m}\mathbf{t}_{2}},\cdots,e^{-j2\pi f_{m}\mathbf{t}_{N_\mathrm{RF}}}]
\right)$ where $\mathbf{t}_{l}\in\mathcal{C}^{K\times 1}=[t_{l,1},t_{l,2},\cdots,t_{l,K}]^{T}$ denotes the time delays realized by $K$ TTDs for the $l$-th path component. 

\begin{figure}
	\centering
	\includegraphics[width=0.47\textwidth]{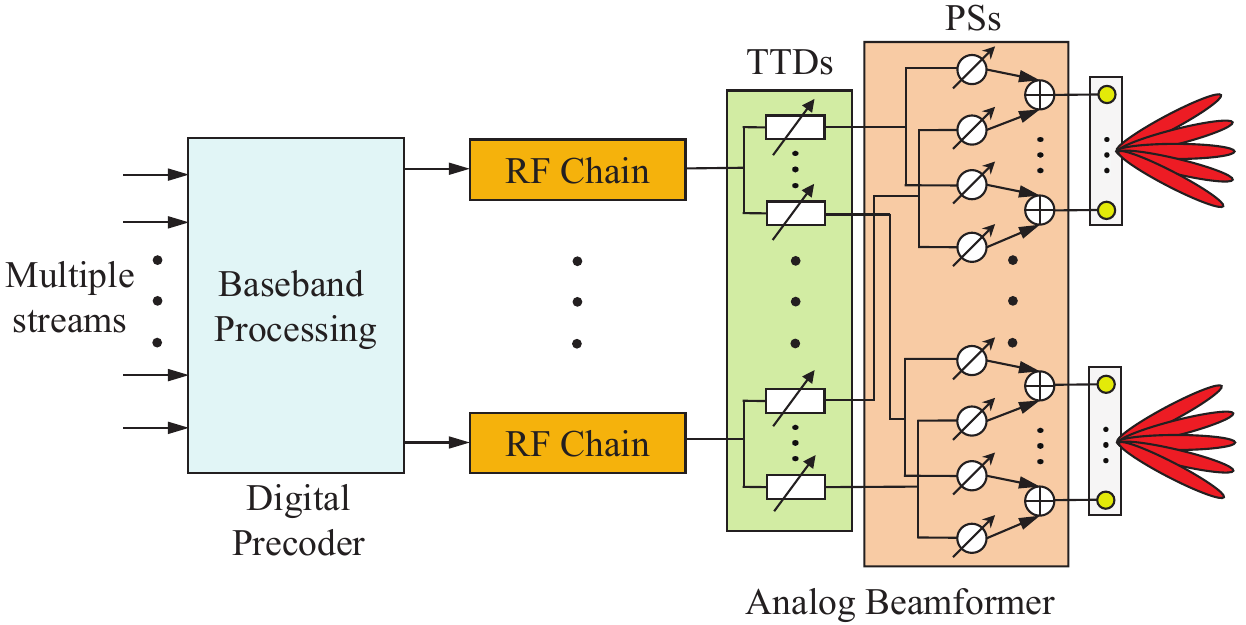}
	\caption{The proposed TTD-DPP structure.}
\end{figure}

Based on the notation above, the beamforming vector for the $l$-th path component $\mathbf{a}_{l,m}=\left[\mathbf{A}_\mathrm{u}\mathbf{A}_{m}^{\mathrm{TTD}}\right]_{[l,:]}=\mathbf{A}_{\mathrm{u},l}e^{-j2\pi f_{m}\mathbf{t}_{l}}=\mathrm{diag}([\bar{\mathbf{a}}_{l,1},\bar{\mathbf{a}}_{l,1},\cdots,\bar{\mathbf{a}}_{l,K}])e^{-j2\pi f_{m}\mathbf{t}_{l}}$. Recalling (\ref{13-1}),  (\ref{14}), (\ref{15}), and \textbf{Lemma 2} in subsection \ref{subBeam}, to eliminate the beam split effect for the $l$-th path component, the phase shifts provided by PSs and the time delays realized by $K$ TTDs have
\begin{equation}\label{34}
[\bar{\mathbf{a}}_{l,1}^{T},\bar{\mathbf{a}}_{l,2}^{T},\cdots,\bar{\mathbf{a}}_{l,K}^{T}]^{T}=\mathbf{f}_\mathrm{t}(\theta_{l}),
\end{equation}
\begin{equation}\label{35}
e^{-j2\pi f_{m}\mathbf{t}_{l}}=\left[1,e^{-j\pi\beta_{l,m}},e^{-j\pi 2\beta_{l,m}},\cdots,e^{-j\pi(K-1)\beta_{l,m}}\right]^{T},
\end{equation}
where the direction rotation factor is $\beta_{l,m}=(\xi_{m}-1)P\theta_{l}$ according to (\ref{23}). Therefore, noting that the value difference between phase shifts of adjacent TTDs is equal, the time delay vector $\mathbf{t}_{l}$ should satisfies the following form as $\mathbf{t}_{l}=[0,s_{l}T_{c},2s_{l}T_{c},\cdots,(K-1)s_{l}T_{c}]^{T}$,
where $T_{c}$ is the period of the carrier frequency $f_\mathrm{c}$, and $s_{l}$ denotes the number of periods that should be delayed for the $l$-th path component. Thus, $s_{l}$ should satisfy
\begin{equation}\label{37}
-2\pi f_{m}s_{l}T_{c}=-\pi\beta_{l,m}.
\end{equation}
Then, substituting $T_{c}=\frac{1}{f_\mathrm{c}}$, $\xi_{m}=\frac{f_{m}}{f_\mathrm{c}}$ and $\beta_{l,m}=(\xi_{m}-1)P\theta_{l}$ in (\ref{23}) into (\ref{37}), we have
\begin{equation}\label{38}
s_{l}=\frac{(\xi_{m}-1)P\theta_{l}}{2\xi_{m}}.
\end{equation}
Note that in (\ref{38}), the number of periods $s_{l}$ is not only decided by the fixed $P$ and the target physical direction $\theta_{l}$, but also decided by the variable relative frequency $\xi_{m}$. This makes (\ref{38}) hard to realize for all $M$ subcarriers, since $s_{l}$ must be fixed due to the hardware constraint of TTDs. To solve this problem, we divide the phase shift $-\pi\beta_{l,m}=-\pi(\xi_{m}-1)P\theta_{l}$ into two parts $-\pi\xi_{m}P\theta_{l}$ and $\pi P\theta_{l}$. The first part is frequency-dependent and can be realized by TTDs with
\begin{equation}\label{39}
s_{l}=P\theta_{l}/{2}.
\end{equation}
Then, the second part $\pi P\theta_{l}$  is frequency-independent, which can be realized through PSs by adding an extra phase shift. Specifically, the phase shifts provided by PSs $\bar{\mathbf{a}}_{l,k},k=1,2,\cdots,K$ should be changed from (\ref{34}) to
\begin{equation}\label{40}
\begin{aligned}
\big[\bar{\mathbf{a}}_{l,1}^{T},\cdots,\bar{\mathbf{a}}_{l,K}^{T}\big]^{T}&=\big[\mathbf{f}_{t}(\theta_{l})_{[1:P]},e^{j\pi P\theta_{l}}\mathbf{f}_{t}(\theta_{l})_{[P+1:2P]},\\
&\cdots,e^{j\pi(K-1)P\theta_{l}}\mathbf{f}_\mathrm{t}(\theta_{l})_{[(K-1)P+1:N_\mathrm{t}]}\big].
\end{aligned}
\end{equation}
Considering the time delays $t_{l,i}$ should be larger than $0$, a small modification is required to be operated on (\ref{39}). Finally, the time delay of the $i$-th delayer $t_{l,i}$ should be 
\begin{equation}\label{41}
t_{l,i}=\left\{
\begin{aligned}
&(K-1)\left|\frac{P\theta_{l}}{2}\right| T_{c}+i\frac{P\theta_{l}}{2} T_{c},\ \theta_{l}<0, \\
&i\frac{P\theta_{l}}{2} T_{c},\ \theta_{l}\geq0,
\end{aligned}
\right.
\end{equation}

 We can observe from (\ref{41}) that the value range of $t_{l,i}$ is $t_{l,i}\in[0,\frac{N_\mathrm{t}}{2}T_{c}]$ with $\theta_{l}\in[-1,1]$ and $1\leq i\leq K$. For example, when $f_\mathrm{c}=300$ GHz, $N_\mathrm{t}=256$, and $K=16$, the range of time delays provided by the TTDs is between $0$ and $426$ ps. It should be noted that several efficient TTDs could meet the time delay requirement\cite{Ref:TTD_Line2018,Ref:TTD2015,Ref:NovTTD2020,Ref:TTD_Filter2017}. For example, the TTD designed based on artificial transmission lines can realize the maximum time delay of $508$ ps with a $4$ ps time delay step, and it can support a $20$ GHz bandwidth \cite{Ref:TTD_Line2018}. Besides, \cite{Ref:TTD2015} proposed a delay-line based TTD, which can realize the maximum time delay of $400$ ps with a $5$ ps time delay step and a $20$ GHz bandwidth. These TTDs are able to support the practical hardware implementation of the proposed TTD-DPP.
 
\begin{algorithm}[htb]
	\caption{Precoding for TTD-DPP.}
	\label{alg:Framwork}
	\begin{algorithmic}[1]
		\REQUIRE ~~\\
		Channel $\mathbf{H}_{m}$; Physical direction $\theta_{l}$
		\ENSURE ~~\\
		Hybrid precoder $\mathbf{A}_\mathrm{u}$, $\mathbf{A}_{m}^\mathrm{TTD}$, and $\mathbf{D}_{m}$		
		\FOR{$l\in\{1,2,\cdots,N_\mathrm{RF}\}$}
		\STATE  $\begin{aligned}
		\big[\bar{\mathbf{a}}_{l,1}^{T},\cdots,\bar{\mathbf{a}}_{l,K}^{T}\big]^{T}&=\big[\mathbf{f}_{t}(\theta_{l})_{[1:P]},e^{j\pi P\theta_{l}}\mathbf{f}_{t}(\theta_{l})_{[P+1:2P]},\\
		&\cdots,e^{j\pi(K-1)P\theta_{l}}\mathbf{f}_\mathrm{t}(\theta_{l})_{[(K-1)P+1:N_\mathrm{t}]}\big]
		\end{aligned}$\\	
		\STATE $\mathbf{A}_{\mathrm{u},l}=\mathrm{blkdiag}([\bar{\mathbf{a}}_{l,1},\bar{\mathbf{a}}_{l,1},\cdots,\bar{\mathbf{a}}_{l,K}])$\\	
		\STATE $t_{l,i}=\left\{
		\begin{aligned}
		&(K-1)\left|\frac{P\theta_{l}}{2}\right| T_{c}+i\frac{P\theta_{l}}{2} T_{c},\ \theta_{l}<0, \\
		&i\frac{P\theta_{l}}{2} T_{c},\ \theta_{l}\geq0,
		\end{aligned}\right.$\\
		\STATE $\mathbf{t}_{l}=[t_{l,1},t_{l,2},\cdots,t_{l,K}]$\\
		\ENDFOR
		\STATE $\mathbf{A}_{u}=[\mathbf{A}_{u,1},\mathbf{A}_{u,2},\cdots,\mathbf{A}_{u,N_\mathrm{RF}}]$
		\FOR{$m\in\{1,2,\cdots,M\}$}
		\STATE $\mathbf{A}_{m}^\mathrm{TTD}=\mathrm{blkdiag}\bigg([
		e^{-j2\pi f_{m}\mathbf{t}_{1}}, \cdots, e^{-j2\pi f_{m}\mathbf{t}_{N_\mathrm{RF}}}]\bigg)$	
		\STATE $\mathbf{H}_{m,\mathrm{eq}}=\mathbf{H}_{m}^{H}\mathbf{A}_{u}\mathbf{A}_{m}^\mathrm{TTD}$
		\STATE $\mathbf{D}_{m}=\mu\mathbf{V}_{m,\mathrm{eq},[:,1:N_\mathrm{RF}]},\mathbf{H}_{m,\mathrm{eq}}=\mathbf{U}_{m,\mathrm{eq}}\mathbf{\Sigma}_{m,\mathrm{eq}}\mathbf{V}_{m,\mathrm{eq}}^{H}$
		\ENDFOR
		\RETURN $\mathbf{A}_{u}$, $\mathbf{A}_{m}^\mathrm{TTD}$ and $\mathbf{D}_{m}$
	\end{algorithmic}
	
\end{algorithm}

By setting $\bar{\mathbf{a}}_{l,i}$ and $t_{l,i}$ as (\ref{40}) and (\ref{41}), the TTD-DPP structure can compensate for the array gain loss caused by the beam split effect for the $l$-th path component. Based on the derivation above, we propose a precoding algorithm for the TTD-DPP structure, where the key idea is to generate beams towards different physical directions of path components at first, and then the time delays are calculated accordingly to make the beam aligned with the physical direction at each subcarrier. Specifically, the pseudo-code is shown in \textbf{Algorithm 1}. At first, for each path component, the analog beamformer $\mathbf{A}_{\mathrm{u},l}$ is calculated in steps $2$-$3$. Then, the time delays that should be delayed by $K$ TTDs are generated in steps $4$-$5$. After that, for each subcarrier frequency, the analog beamformer $\mathbf{A}_{m}^\mathrm{TTD}$ is generated in step $9$. Finally, the digital precoder $\mathbf{D}_{m}$ is also calculated based on singular value decomposition (SVD) precoding in steps $10$ and $11$, where $\mu$ is the power normalization coefficient. By utilizing the \textbf{Algorithm 1}, the TTD-DPP can realize the near-optimal achievable rate, which will be verified by simulations in Section \ref{Sim}.

It should be emphasized that we only provide one practical hardware implementation of DPP in this paper. Actually, any hardware component that can realize frequency-dependent phase shifts is able to realize the concept of DPP. For instance, multiple RF chains can be used to realize frequency-dependent phase shifts in the baseband. Specifically, $N_\mathrm{RF}$ RF chain group with each group containing $K$ RF chains can be utilized, where each RF chain group connects to all antenna elements through PSs in a sub-connected manner. In this way, when the PSs generate frequency-independent beams according to (\ref{14}) and the baseband signal processing realizes the frequency-dependent phase shifts $\beta_{m}$ according to (\ref{23}), the mechanism of the DPP proved in \textbf{Lemma 2} can also be realized, and thus the beam split effect can be eliminated.

In addition, without loss of generality, we consider the single-user scenario to provide an insightful explanation of the beam split effect and the proposed TTD-DPP structure in this paper. For the multi-user scenario, the proposed TTD-DPP can also achieve the near-optimal performance by simply modifying existing multi-user precoding algorithms, e.g., following the idea of beam selection based precoding\cite{Ref:NObeamSe2016}. To be more specific, the TTDs and PSs generate beams aligned with different users by following the design principle in \textbf{Lemma 2}, then the digital precoder can be designed by utilizing zero-forcing or block diagonalization based precoding method. Since the proposed TTD-DPP can realize the near-optimal array gain across the entire bandwidth, the near-optimal achievable rate can be expected in the multi-user scenario\cite{Ref:NObeamSe2016}.

\subsection{Achievable rate performance}\label{RateAnalysis}
In this subsection, we will derive the achievable rate of the proposed TTD-DPP structure. The achievable rate $R$ of an $M$-subcarrier wideband THz massive MIMO system is\cite{Ref:FundWC2005}
\begin{equation}\label{42}
\begin{aligned}
R&=\sum_{m=1}^{M}R_{m}\\
&=\sum_{m=1}^{M}\log_{2}{\Bigg(\left|\mathbf{I}_{N_\mathrm{r}}+\frac{\rho}{N_\mathrm{s}\sigma^{2}}\mathbf{H}_{m}\mathbf{A}_{m}\mathbf{D}_{m}\mathbf{D}_{m}^{H}\mathbf{A}_{m}^{H}\mathbf{H}_{m}^{H}\right|\Bigg)},
\end{aligned}
\end{equation}
where $R_{m}$ is the achievable rate at the $m$-th subcarrier, and $\mathbf{A}_{m}=\mathbf{A}_\mathrm{u}\mathbf{A}_{m}^\mathrm{TTD}$ in TTD-DPP. By utilizing the ordered SVD of $\mathbf{H}_{m}$ as $\mathbf{H}_{m}=\mathbf{U}_{m}\mathbf{\Sigma}_{m}^\mathrm{f}\mathbf{V}_{m}^{H}$, $R_{m}$ can be converted into \cite{Ref:SpatiallyPre2014}
\begin{equation}\label{43}
R_{m}=\log_{2}{\Bigg(\left|\mathbf{I}_{d_\mathrm{m}}+\frac{\rho}{N_\mathrm{s}\sigma^{2}}{\mathbf{\Sigma}_{m}^\mathrm{f}}^{2}{\mathbf{V}_{m}^\mathrm{f}}^{H}\mathbf{A}_{m}\mathbf{D}_{m}\mathbf{D}_{m}^{H}\mathbf{A}_{m}^{H}\mathbf{V}_{m}^\mathrm{f}\right|\Bigg)},
\end{equation}
where the diagonal matrix $\mathbf{\Sigma}_{m}^\mathrm{f}=\mathrm{diag}([\lambda_{1},\lambda_{2},\cdots,\lambda_{d_{m}}])\in\mathcal{C}^{d_{m}\times d_{m}}$ ($\lambda_{i},i=1,2,\cdots,d_{m}$) representing the singular value of $\mathbf{H}_{m}$, and the matrix $\mathbf{V}_{m}^\mathrm{f}\in\mathcal{C}^{N_\mathrm{t}\times d_{m}}$ with ${\mathbf{V}_{m}^\mathrm{f}}^{H}\mathbf{V}_{m}^\mathrm{f}=\mathbf{I}_{d_{m}}$ are obtained from the ordered SVD of the channel $\mathbf{H}_{m}$, where $d_{m}$ denotes the rank of $\mathbf{H}_{m}$. 

We assume that the parameters $(N_\mathrm{t},N_\mathrm{RF},N_\mathrm{s})$ for DPP are designed to fully exploit the multiplexing gain from the multi-path channel. This assumption can be easily satisfied in practical THz massive MIMO systems\cite{Ref:SpatiallyPre2014}. Hence, according to \cite{Ref:SpatiallyPre2014}, (\ref{43}) can be rewritten as
\begin{equation}\label{44}
R_{m}=\log_{2}{\Bigg(\left|\mathbf{I}_{N_{s}}+\frac{\rho}{N_{s}\sigma^{2}}\mathbf{\Sigma}_{m}^{2}{\mathbf{V}_{m}}^{H}\mathbf{A}_{m}\mathbf{D}_{m}\mathbf{D}_{m}^{H}\mathbf{A}_{m}^{H}\mathbf{V}_{m}\right|\Bigg)},
\end{equation}
where $\mathbf{\Sigma}_m=\mathbf{\Sigma}_{m,[1:N_{s},1:N_{s}]}^\mathrm{f}$ and $\mathbf{V}_{m}=\mathbf{V}_{m,[:,1:N_{s}]}^\mathrm{f}$. Note that $\mathbf{V}_{m}$ is the optimal precoding matrix without any hardware constraint. Since the steering vectors $\mathbf{f}_\mathrm{t}(\bar{\theta}_{l,m})$ in (\ref{3}) are approximately orthogonal due to the large number of antennas \cite{Ref:CFAoD2018}, these vectors $\mathbf{f}_\mathrm{t}(\bar{\theta}_{l,m})$ can form a set of orthogonal basis of $\mathbf{H}\mathbf{H}^{H}$. Notice that the columns of $\mathbf{V}_{m}^\mathrm{f}$ are also a set of orthogonal basis for $\mathbf{H}\mathbf{H}^{H}$, the columns of $\mathbf{V}_{m}$ can be approximately seen as the linear combination of $\mathbf{f}_\mathrm{t}(\bar{\theta}_{l,m})$ as
\begin{equation}\label{45}
\mathbf{V}_{m}\approx\mathbf{A}_\mathrm{t}\mathbf{D}_{m,\mathrm{opt}},
\end{equation}
where $\mathbf{A}_\mathrm{t}=[\mathbf{f}_\mathrm{t}(\bar{\theta}_{1,m}),\mathbf{f}_\mathrm{t}(\bar{\theta}_{2,m}),\cdots,\mathbf{f}_\mathrm{t}(\bar{\theta}_{N_\mathrm{RF},m})]$ with $\bar{\theta}_{l,m}$ being sorted by path gains $|g_{1}|>|g_{2}|>\cdots>|g_{N_\mathrm{RF}}|$, and $\mathbf{D}_{m,\mathrm{opt}}\in\mathcal{C}^{N_\mathrm{RF}\times N_\mathrm{s}}$. Note that in (\ref{45}), $\mathbf{V}_{m}$ and $\mathbf{A}_\mathrm{t}$ satisfy ${\mathbf{V}_{m}}^{H}\mathbf{V}_{m}=\mathbf{I}_{N_\mathrm{s}}$ and $\mathbf{A}_\mathrm{t}^{H}\mathbf{A}_\mathrm{t}=\mathbf{I}_{N_\mathrm{RF}}$. Therefore, the optimal digital precoder $\mathbf{D}_{m,\mathrm{op}}$ also satisfies $\mathbf{D}_{m,\mathrm{opt}}^{H}\mathbf{D}_{m,\mathrm{opt}}=\mathbf{I}_{N_\mathrm{s}}$. Obviously, when $\mathbf{A}_{m}=\mathbf{A}_\mathrm{t}$ and $\mathbf{D}_{m}=\mathbf{D}_{m,\mathrm{opt}}$, the optimal achievable rate $R_{m,\mathrm{opt}}$ is $R_{m,\mathrm{opt}}=\log_{2}{\left(\left|\mathbf{I}_{N_\mathrm{s}}+\frac{\rho}{N_\mathrm{s}\sigma^{2}}\mathbf{\Sigma}_{m}^{2}\right|\right)}$.

However, for the proposed TTD-DPP, the near-optimal precoder $\mathbf{A}_{m}=\mathbf{A}_\mathrm{t}$ and $\mathbf{D}_{m}=\mathbf{D}_{m,\mathrm{opt}}$ cannot be achieved at all subcarriers, since the analog beamforming vectors $\mathbf{a}_{l,m}$ as shown in (\ref{13-1}) in DPP cannot be equal to $\mathbf{f}_\mathrm{t}(\bar{\theta}_{l,m})$ at all subcarriers. We can obtain the achievable rate $R_{m,\mathrm{TTD}}$ of the proposed TTD-DPP by substituting (\ref{45}) into (\ref{44}) as
\begin{equation}\label{47}
R_{m,\mathrm{TTD}}=\log_{2}{\Bigg(\left|\mathbf{I}_{N_\mathrm{s}}+\frac{\rho}{N_{s}\sigma^{2}}\mathbf{\Sigma}_{m}^{2}\mathbf{V}_{m,\mathrm{eq}}^{H}\mathbf{V}_{m,\mathrm{eq}}\right|\Bigg)},
\end{equation}
where   $\mathbf{V}_{m,\mathrm{eq}}=\mathbf{D}_{m}^{H}\mathbf{A}_{m}^{H}\mathbf{A}_\mathrm{t}\mathbf{D}_{m,\mathrm{opt}}$. Considering that the columns of $\mathbf{A}_{m}$ are the analog beamforming vectors generated by the DPP, which can generate beams aligned with the physical direction $\theta_{l}$, and the steering vectors towards different physical directions are approximately orthogonal\cite{Ref:CFAoD2018}, we have $\mathbf{A}_{m}^{H}\mathbf{A}_\mathrm{t}=\mathrm{blkdiag}\big(\big[\mathbf{a}_{1,m}^{H}\mathbf{f}_\mathrm{t}(\bar{\theta}_{1,m}),\mathbf{a}_{2,m}^{H}\mathbf{f}_\mathrm{t}(\bar{\theta}_{2,m}),\cdots,\mathbf{a}_{N_\mathrm{RF},m}^{H}\mathbf{f}_\mathrm{t}(\bar{\theta}_{N_\mathrm{RF},m})\big]\big)$. Recalling the array gain $\eta(\mathbf{a}_{l,m},\theta_{l},f_{m})=|\mathbf{f}_\mathrm{t}(2d\frac{f_{m}}{c}\theta_{l})^{H}{\mathbf{a}}_{l,m}|$, we know that $\mathbf{\Phi}_{m}=\mathbf{A}_{m}^{H}\mathbf{A}_\mathrm{t}$ satisfies
\begin{equation}\label{49}
\begin{aligned}
\mathbf{\Phi}_{m}^{H}\mathbf{\Phi}_{m}=&\mathrm{blkdiag}\big(\big[\eta(\mathbf{a}_{1,m},\theta_{1},f_{m})^{2},\eta(\mathbf{a}_{2,m},\theta_{2},f_{m})^{2},\\
&\cdots,\eta(\mathbf{a}_{N_\mathrm{RF},m},\theta_{N_\mathrm{RF}},f_{m})^{2}\big]\big).
\end{aligned}
\end{equation}
Therefore, based on (\ref{49}), the achievable rate $R_{m,\mathrm{TTD}}$ becomes
\begin{equation}\label{50}
\begin{aligned}
R&_{m,\mathrm{TTD}}\\
&=\log_{2}{\Bigg(\left|\mathbf{I}_{N_\mathrm{s}}+\frac{\rho\mathbf{\Sigma}_{m}^{2}\mathbf{D}_{m,\mathrm{opt}}^{H}\mathbf{\Phi}_{m}^{H}\mathbf{D}_{m}\mathbf{D}_{m}^{H}\mathbf{\Phi}_{m}\mathbf{D}_{m,\mathrm{opt}}}{N_\mathrm{s}\sigma^{2}}\right|\Bigg)}\\
&\overset{(a)}{\approx}\log_{2}{\Bigg(\left|\mathbf{I}_{N_\mathrm{s}}+\frac{\rho\mathbb{E}(\eta(\mathbf{a}_{l,m},\theta_{l},f_{m})^{2})\mathbf{\Sigma}_{m}^{2}\mathbf{D}_{m,\mathrm{opt}}^{H}\mathbf{D}_{m,\mathrm{opt}}}{N_\mathrm{s}\sigma^{2}}\right|\Bigg)}\\
&\overset{(b)}{=}\log_{2}{\left(\left|\mathbf{I}_{N_\mathrm{s}}+\frac{\rho}{N_\mathrm{s}\sigma^{2}}\mathbb{E}(\eta(\mathbf{a}_{l,m},\theta_{l},f_{m}))^{2}\mathbf{\Sigma}_{m}^{2}\right|\right)},
\end{aligned}
\end{equation}
where $(a)$ is achieved by setting $\mathbf{D}_{m}=\mathbf{D}_{m,\mathrm{opt}}$, $\mathbf{D}_{m,\mathrm{opt}}^{H}\mathbf{D}_{m,\mathrm{opt}}=\mathbf{I}_{N_\mathrm{s}}$ and the assumption $\eta(\mathbf{a}_{1,m},\theta_{1},f_{m})^{2}\approx\eta(\mathbf{a}_{2,m},\theta_{2},f_{m})^{2}\approx\cdots\approx\eta(\mathbf{a}_{N_\mathrm{RF},m},\theta_{N_\mathrm{RF}},f_{m})^{2}\approx\mathbb{E}(\eta(\mathbf{a}_{l,m},\theta_{l},f_{m})^{2})$ which is reasonable since the beams generated by the DPP can achieve near-optimal array gain across the entire bandwidth, and $(b)$ comes from $\mathbf{D}_{m,\mathrm{opt}}^{H}\mathbf{D}_{m,\mathrm{opt}}=\mathbf{I}_{N_\mathrm{s}}$.

We can observe from (\ref{50}) that the achievable rate $R_{m,\mathrm{TTD}}$ is mainly decided by the array gain obtained in different physical directions at different subcarriers. This indicates that the array gain loss caused by the beam split effect is vital to the achievable rate. Based on (\ref{50}), the ratio between the achievable rate achieved by the TTD-DPP and the optimal achievable rate has
\begin{equation}\label{51}
\begin{aligned}
\frac{R_{m,\mathrm{TTD}}}{R_{m,\mathrm{opt}}}&=\frac{\log_{2}{\left(\left|\mathbf{I}_{N_\mathrm{s}}+\frac{\rho}{N_\mathrm{s}\sigma^{2}}\mathbb{E}(\eta(\mathbf{a}_{l,m},\theta_{l},f_{m}))^{2}\mathbf{\Sigma}_{m}^{2}\right|\right)}}{\log_{2}{\left(\left|\mathbf{I}_{N_\mathrm{s}}+\frac{\rho}{N_\mathrm{s}\sigma^{2}}\mathbf{\Sigma}_{m}^{2}\right|\right)}}\\
&=\frac{\Sigma_{l=1}^{N_\mathrm{s}}\log_{2}{(1+\frac{\rho}{N_\mathrm{s}\sigma^{2}}\mathbb{E}({\eta(\mathbf{a}_{l,m},\theta_{l},f_{m})}^{2})\lambda_{l}^{2})}}{\Sigma_{l=1}^{N_\mathrm{s}}\log_{2}{(1+\frac{\rho}{N_\mathrm{s}\sigma^{2}}\lambda_{l}^{2})}}\\
&\overset{(a)}{>}\frac{\Sigma_{l=1}^{N_\mathrm{s}}\frac{\rho}{N_\mathrm{s}\sigma^{2}}\mathbb{E}({\eta(\mathbf{a}_{l,m},\theta_{l},f_{m})}^{2})\lambda_{l}^{2}}{\Sigma_{l=1}^{N_\mathrm{s}}\frac{\rho}{N_\mathrm{s}\sigma^{2}}\lambda_{l}^{2}},
\end{aligned}
\end{equation}
where $(a)$ is based on $\log_{2}(1+x)<x$ and ${\eta(\mathbf{a}_{l,m},\theta_{l},f_{m})}^{2}\leq1$. Then, we have
\begin{equation}\label{52}
\begin{aligned}
\frac{R_{m,\mathrm{TTD}}}{R_{m,\mathrm{opt}}}&>\frac{\Sigma_{l=1}^{N_\mathrm{RF}}\frac{\rho}{N_\mathrm{s}\sigma^{2}}{\mathbb{E}(\eta(\mathbf{a}_{l,m},\theta_{l},f_{m})}^{2})\lambda_{l}^{2}}{\Sigma_{l=1}^{N_\mathrm{RF}}\frac{\rho}{N_\mathrm{s}\sigma^{2}}\lambda_{l}^{2}}\\
&={\mathbb{E}\Big(\eta(\mathbf{a}_{l,m},\theta_{l},f_{m})^{2}\Big)}.
\end{aligned}
\end{equation}
Similar to the process to compute $\mathbb{E}\left(\eta(\mathbf{a}_{l,m},\theta_{l},f_{m})\right)$ in (\ref{29}), the polynomial fitting with three points $(-1,\Xi_{P}(1-\xi_{m})^{2})$, $(0,P^{2})$ and $(1,\Xi_{P}(\xi_{m}-1)^{2})$ is utilized to calculate ${\mathbb{E}(\eta(\mathbf{a}_{l,m},\theta_{l},f_{m})^{2})}$ in (\ref{52}) as
\begin{equation}\label{53}
\begin{aligned}
\mathbb{E}&\Big(\eta(\mathbf{a}_{l,m},\theta_{l},f_{m})^{2}\Big)=\frac{K^{2}}{2MN_\mathrm{t}^{2}}\sum_{m=1}^{M}\int_{-1}^{1}\left[\Xi_{P}((\xi_{m}-1)\theta_{l})\right]^{2}\mathrm{d}\theta_{l}\\
&\approx\frac{K^{2}}{2MN_\mathrm{t}^{2}}\sum_{m=1}^{M}\int_{-1}^{1}\left[(\Xi_{P}(\xi_{m}-1)^{2}-P^{2})\theta_{l}^{2}+P^{2}\right]\mathrm{d}\theta_{l}\\
&=\frac{K^{2}}{MN_\mathrm{t}^{2}}\sum_{m=1}^{M}\frac{1}{3}\Xi_{P}(\xi_{m}-1)^{2}+\frac{2}{3}P^{2}.
\end{aligned}
\end{equation}
It is clear from (\ref{53}) that the achievable rate $R_{m,\mathrm{TTD}}$ of the TTD-DPP structure will increase as $K$ increases, since the wider mainlobe of $\Xi_{P}(x)$ results in the fact that $|\Xi_{P}(\xi_{m}-1)|$ is closer to $1$. While, a small number of $K$ is usually enough to realize the near-optimal achievable rate performance. For instance, when $f_\mathrm{c}=300$ GHz, $B=15$ GHz, $M=128$, $K=8$ and $N_\mathrm{t}=256$, we have $\frac{R_{m,\mathrm{TTD}}}{R_{m,\mathrm{opt}}}>{\mathbb{E}(\eta(\mathbf{a}_{l,m},\theta_{l},f_{m})^{2})}=0.94$. This means that by adopting the proposed TTD-DPP, which can efficiently mitigate the beam split effect, the near-optimal achievable rate can be achieved.
\vspace{-3mm}
\section{Simulation Results}\label{Sim}

\begin{table}[bp]
	\centering
	\caption{System Parameters for Simulations}\label{tab:tab2}
	\renewcommand\arraystretch{1.25}
	\begin{tabular}{|c|c|}
		\hline
		The number of the BS antennas $N_{\mathrm{t}}$ & $256$ \\\hline
		The number of the user antennas $N_{\mathrm{r}}$ & $1,2,4$ \\\hline
		The number of channel paths $L$ & $4$\\\hline
		The central frequency $f_\mathrm{c}$ & $300$ GHz \\\hline
		The bandwidth $B$ & $30$ GHz\\\hline
		The number of the subcarriers $M$ & $128$ \\\hline
		The number of RF chains $N_{\mathrm{RF}}$ & $4$\\\hline
		The number of TD elements  $K$ & $2\sim16$\\\hline
		Physical directions of the paths $\tilde{\theta}_{l},\tilde{\phi}_{l}$ & $\mathcal{U}[-\frac{\pi}{2},\frac{\pi}{2}]$\\\hline
	\end{tabular}
\end{table}

\begin{figure*}
	\centering
	\includegraphics[width=0.92\textwidth]{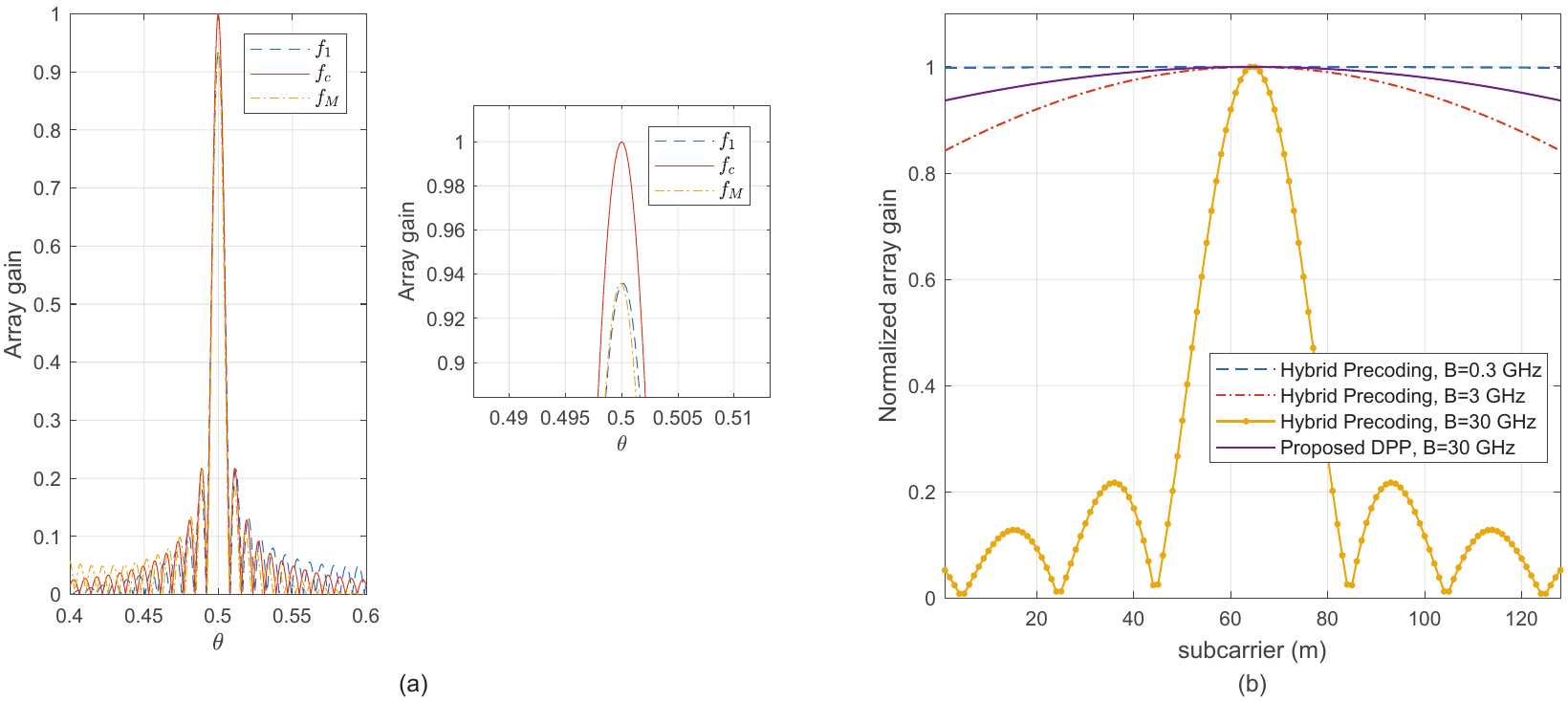}
	\caption{Normalized array gain performance: (a) on different subcarriers; (b) against different subcarriers.}
\end{figure*}

In this section, we provide simulation results to verify the performance of the proposed TTD-DPP to realize the concept of DPP for wideband THz massive MIMO. The main simulation parameters are shown in Table I\footnote{We assume $N_\mathrm{RF}=L$ in simulations, since the number of RF chains is usually larger than or equal to the maximum number of channel paths in practice to guarantee the hybrid precoding performance\cite{Ref:SpatiallyPre2014}.}. The ULAs are considered at the BS and the user. We assume that the number of streams is equal to the number of receive antennas, i.e., $N_\mathrm{s}=N_\mathrm{r}$. The performance of conventional hybrid precoding methods using PSs is also provided for comparison. In the proposed TTD-DPP, the number of TTDs is set as $K=16$ except for Fig. 11 and Fig. 12, and the bandwidth is $30$ GHz except for Fig. 12. We utilize (\ref{42}) to calculate the achievable rate. The transmission signal-to-noise ratio (SNR) is defined as $\rho/\sigma^{2}$.

Fig. 8 shows the normalized array gain performance. Fig. 8 (a) illustrates the normalized array gain against the physical direction of the beamforming vector $\mathbf{a}_{l,1}$ at the minimum subcarrier frequency $f_{1}$, $\mathbf{a}_{l,M}$ at the maximum subcarrier frequency $f_{M}$,  and the beamforming vector at the central frequency $f_\mathrm{c}$. They are generated by the proposed DPP with the target physical direction $\theta_{l}=0.5$. We can observe from Fig. 8 (a) that at the minimum frequency $f_{1}$ and the maximum frequency $f_{M}$, the beamforming vector $\mathbf{a}_{l,1}$ and $\mathbf{a}_{l,M}$ can be aligned with the target physical direction $\theta_{l}$. Thus, we can conclude that by using the proposed DPP, the user can be covered by beams at different subcarrier frequencies, which efficiently mitigates the array gain loss caused by the beam split effect. More than $94\%$ of the optimal array gain can be achieved by the proposed DPP at $f_{1}$ and $f_{M}$. The normalized array gain performance of the proposed DPP at different subcarriers are shown in Fig. 8 (b). The target physical direction is still set as $\theta_{l}=0.5$, and the bandwidth is $B=30$ GHz. The array gains achieved by the conventional hybrid precoding architecture with different bandwidths $B=0.3,\ 3,\ 30$ GHz are also provided for comparison. We can observe that when $B=30$ GHz, the conventional hybrid precoding architecture \cite{Ref:SpatiallyPre2014} suffers from severe array gain loss due to the beam split effect, e.g., $80\%$ array gain loss at most of subcarriers. In contrast, the proposed DPP can realize almost flat array gain across the entire bandwidth.

\begin{figure}
	\centering
	\includegraphics[width=0.47\textwidth]{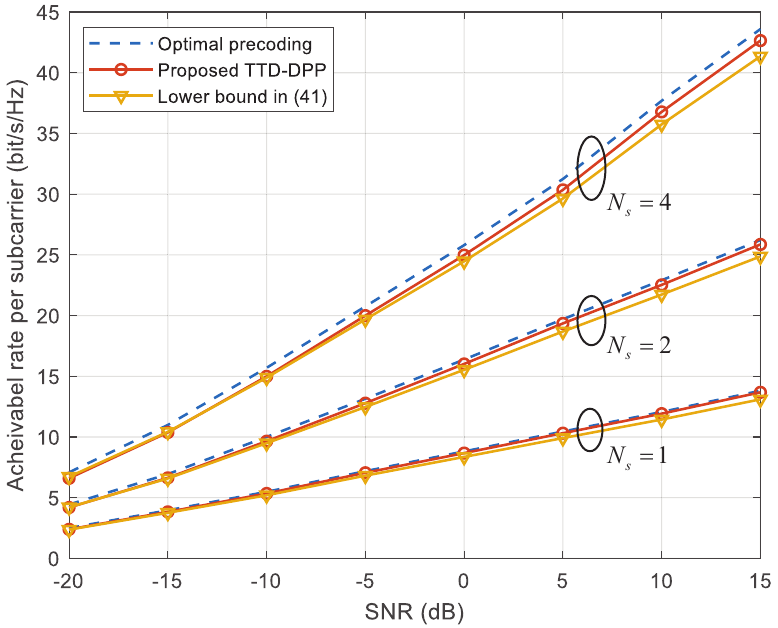}
	\caption{Achievable rate performance versus the transmission SNR for the proposed TTD-DPP.}
\end{figure}

Fig. 9 illustrates the average achievable rate of the proposed TTD-DPP, where different numbers of data streams $N_\mathrm{s}=1,\ 2,\ 4$ are considered. We provide the achievable rate performance of the optimal unconstrained fully-digital precoding \cite{Ref:mmWaveMIMO2016} and the theoretical result in (\ref{53}) as the upper bound and the lower bound, respectively. We can find that the proposed TTD-DPP can achieve more than $95$\% of the optimal achievable rate, and the actual achievable rate is always larger than the lower bound in (\ref{53}), which is consistent with the analysis in the subsection \ref{RateAnalysis}. 

\begin{figure}
	\centering
	\includegraphics[width=0.47\textwidth]{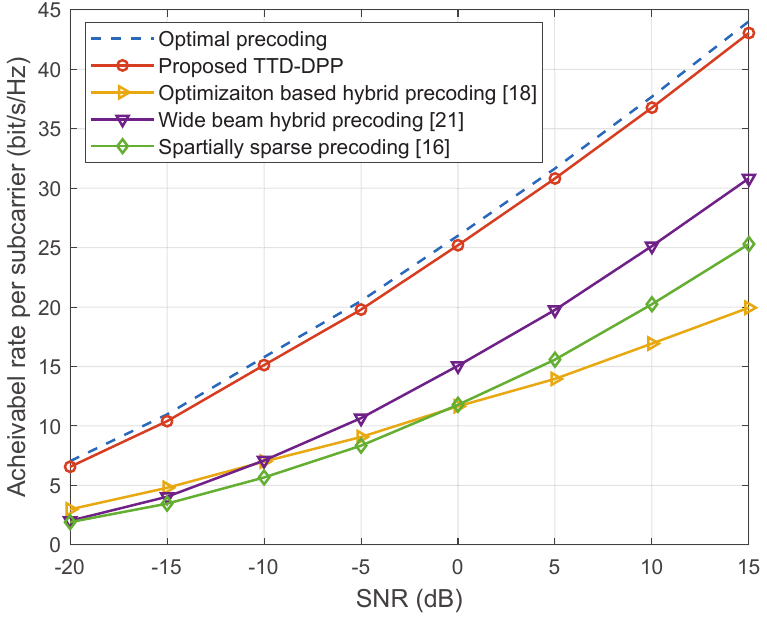}
	\caption{Achievable rate performance comparison between the proposed scheme and existing schemes.}
\end{figure}

Fig. 10 compares the average achievable rate performance between the proposed TTD-DPP and other existing hybrid precoding schemes when $N_\mathrm{s}=4$. The existing solutions include the spatially sparse precoding \cite{Ref:SpatiallyPre2014}, the achievable rate optimization \cite{Ref:SubWideHy2017}, and the wide beam based hybrid precoding \cite{Ref:WideCode2019}. Specifically, we can observe from Fig. 10 that the spatially sparse precoding \cite{Ref:SpatiallyPre2014} suffers a nearly $50\%$ achievable rate loss caused by the beam split effect. Although the achievable rate optimization\cite{Ref:SubWideHy2017} and wide beam based hybrid precoding\cite{Ref:WideCode2019} designed for mmWave massive MIMO systems can partially relieve the achievable rate loss incurred by the beam split effect, the performance is still unacceptable. On the contrary, the proposed TTD-DPP scheme can significantly outperform these existing schemes and can achieve the near-optimal achievable rate, e.g., more than $95$\% of the optimal achievable rate.

\begin{figure}
	\centering
	\includegraphics[width=0.47\textwidth]{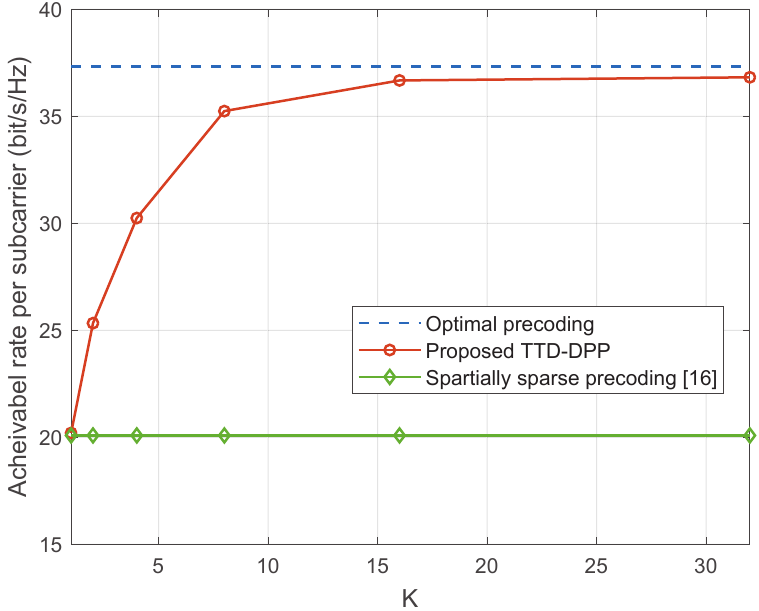}
	\caption{Achievable rate performance versus the number of TTDs $K$.}
\end{figure}

We provide the achievable rate performance of the proposed TTD-DPP versus the number of TTDs $K$ in Fig. 11. The SNR is set as $10$ dB, and $N_\mathrm{r}=4$. We can observe from Fig. 11 that the achievable rate performance of the proposed TTD-DPP increases as the number of TTDs $K$ grows up. It should be noted that the proposed TTD-DPP can realize more than $95$\% of the optimal achievable rate when $K$ is bigger than $16$. This indicates that the required number of TTDs $K$ in the proposed TTD-DPP is much smaller than the number of antennas $N$, e.g., $K=16\ll N=256$. This small number of TTDs guarantees an acceptable power consumption, which will be verified by the energy efficiency comparison later.

\begin{figure}
	\centering
	\includegraphics[width=0.47\textwidth]{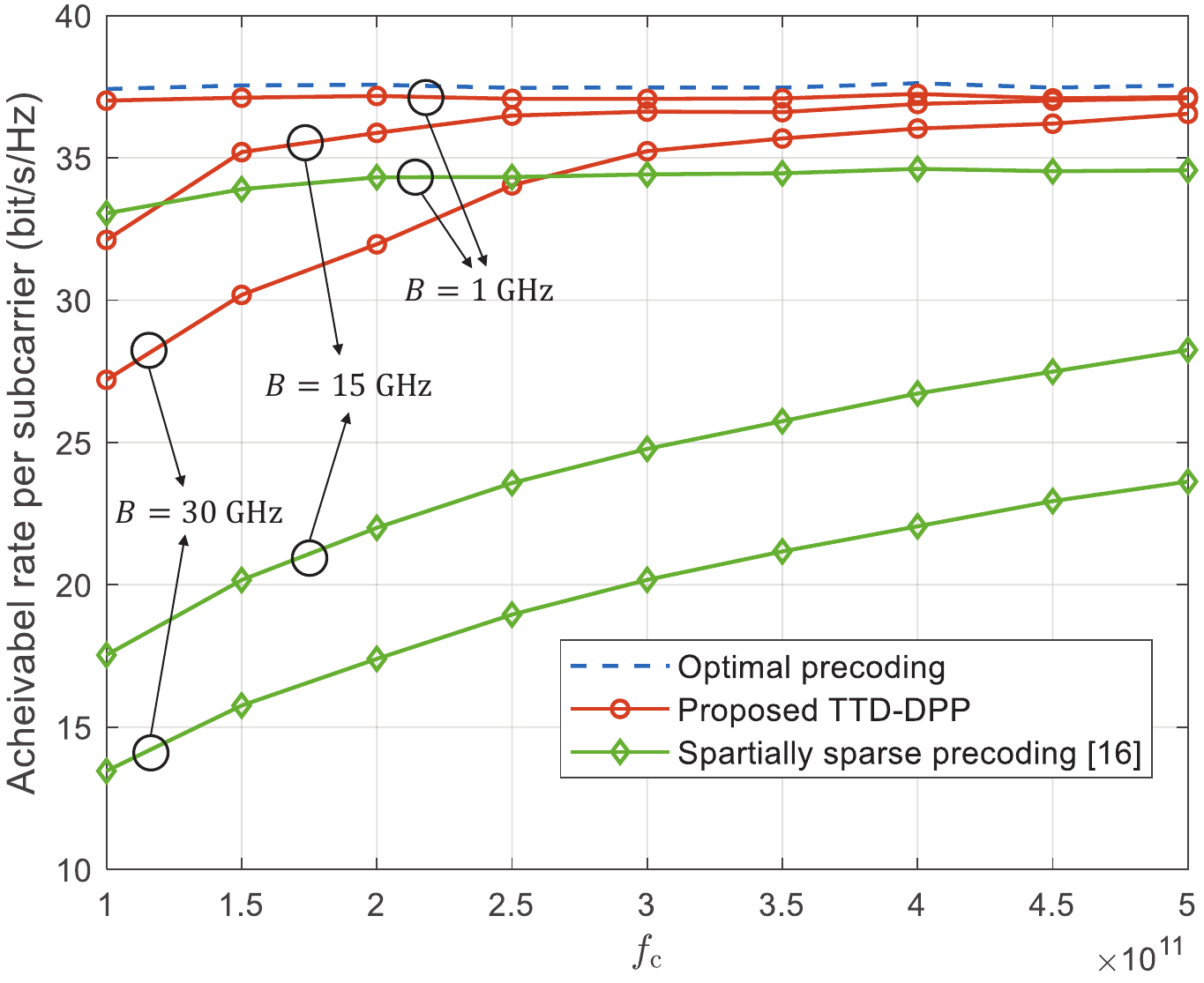}
	\caption{Achievable rate performance versus the central frequency $f_\mathrm{c}$.}
\end{figure}

Fig. 12 shows the achievable rate performance versus the central frequency $f_\mathrm{c}$ with different bandwidths $B=1, 15, 30$ GHz. The SNR is set as 10 dB, and the number of TD elements is $K=8$. We can observe from Fig. 12 that for the fixed number of TD elements $K$ and antenna number $N_\mathrm{t}$, the performance of the proposed TTD-DPP increases when the central frequency $f_\mathrm{c}$ becomes larger or the bandwidth $B$ becomes smaller. Once (19) is satisfied, the proposed TTD-DPP could achieve the near-optimal achievable rate performance. Meanwhile, even when the proposed TTD-DPP cannot completely eliminate the performance loss induced by the beam split effect, its performance is still better than the conventional hybrid precoding \cite{Ref:SpatiallyPre2014}.

\begin{figure*}
	\centering
	\includegraphics[width=0.86\textwidth]{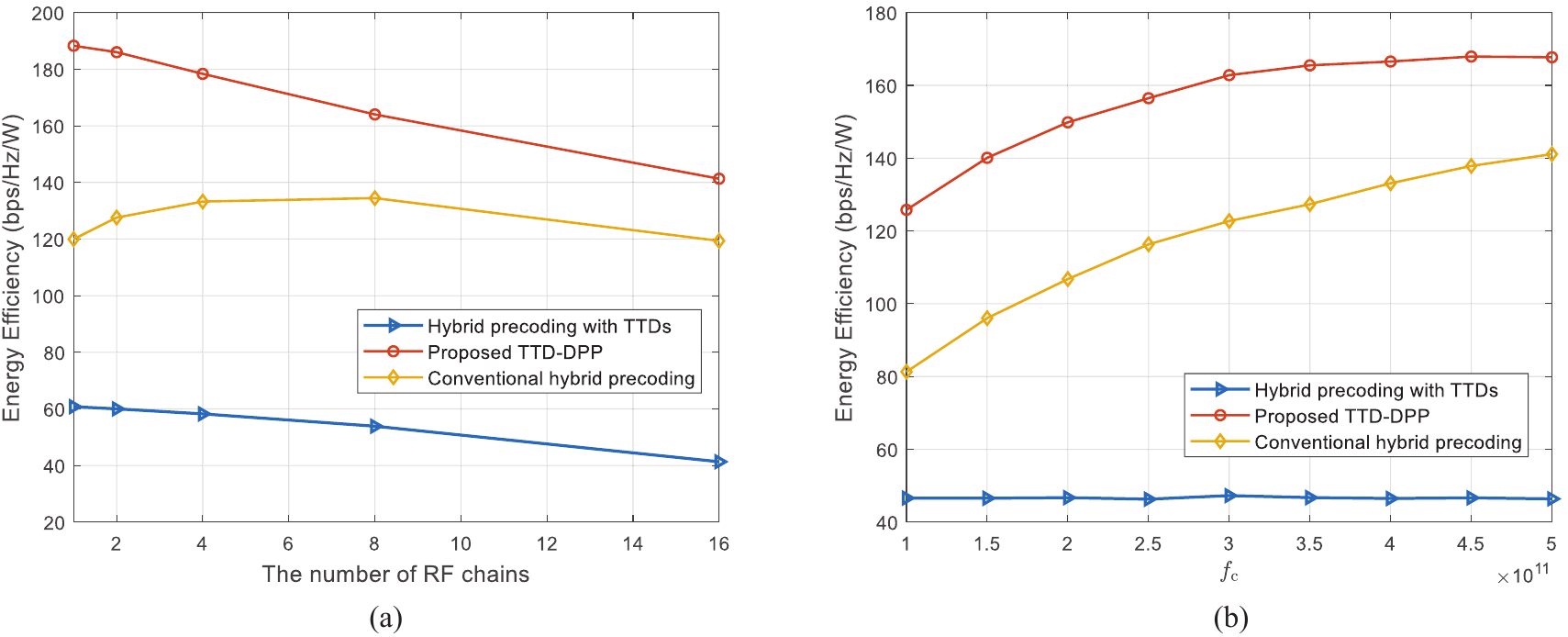}
	\caption{Energy efficiency comparison: (a) Energy efficiency versus the number of RF chains; (b) Energy efficiency versus central frequency.}
\end{figure*}

Finally, Fig. 13 provides the energy efficiency comparison when $N_\mathrm{RF}=N_\mathrm{s}$ varies from $1$ to $16$ in Fig. 13 (a) and $f_\mathrm{c}$ varies with $N_\mathrm{RF}=N_\mathrm{s}=4$ in Fig. 13 (b). The energy efficiency is defined as the ratio between the achievable rate and the power consumption. Specifically, we compare the energy efficiency of the conventional PSs based hybrid precoding architecture\cite{Ref:SpatiallyPre2014}, the TTDs based hybrid precoding where the PSs are replaced by TTDs\cite{Ref:TTD2019}, and the proposed TTD-DPP. The power consumption of these three schemes are denoted as $P_\mathrm{HP},P_\mathrm{TTD},P_\mathrm{DPP}$, respectively. We have
$P_\mathrm{HP}=P_\mathrm{t}+P_\mathrm{BB}+N_\mathrm{RF}P_\mathrm{RF}+N_\mathrm{RF}NP_\mathrm{PS}$,
$P_\mathrm{TTD}=P_\mathrm{t}+P_\mathrm{BB}+N_\mathrm{RF}P_\mathrm{RF}+N_\mathrm{RF}NP_\mathrm{TTD}$, and $P_\mathrm{DPP}=P_\mathrm{t}+P_\mathrm{BB}+N_\mathrm{RF}P_\mathrm{RF}+N_\mathrm{RF}KP_\mathrm{TTD}+N_\mathrm{RF}NP_\mathrm{PS}$,
where $P_\mathrm{t}$ is the transmission power, and $P_\mathrm{BB}$, $P_\mathrm{RF}$, $P_\mathrm{PS}$, and $P_\mathrm{TTD}$ denote the power consumption of baseband processing, RF chain, PS, and TTD, respectively. Here, we adopt the practical values as $\rho=30$ mW\cite{Ref:EnHP2016}, $P_\mathrm{BB}=300$ mW\cite{Ref:PSorSW2016}, $P_\mathrm{RF}=200$ mW\cite{Ref:EnHP2016}, $P_\mathrm{PS}=20$ mW\cite{Ref:PSorSW2016} and $P_\mathrm{TTD}=100$ mW\cite{Ref:TTD_Line2018}. We can observe from Fig. 13 (a) and (b) that the proposed TTD-DPP enjoys higher energy efficiency than the conventional hybrid precoding architecture using PSs and TTDs. Besides, Fig. 13 (b) shows that the energy efficiency of the proposed TTD-DPP decreases when $f_{M}/f_\mathrm{c}$ becomes larger, since the larger $f_{M}/f_\mathrm{c}$ results in the more severe beam split effect. Under this circumstance, the fixed number of TD elements may not be able to fully eliminate the achievable rate loss caused by the beam split effect. However, we can see that the proposed TTD-DPP always has higher energy efficiency than the conventional hybrid precoding architecture with TTDs. This is because much fewer TTDs ($N_\mathrm{RF}K$ instead of $N_\mathrm{RF}N$, where $K\ll N$) are utilized in the TTD-DPP. In conclusion, the energy efficiency performance shows that the proposed TDD-DPP is able to provide a better tradeoff between the achievable rate performance and power consumption.
\vspace{-3mm}

\section{Conclusions}\label{Con}

In this paper, we have investigated wideband precoding for future THz massive MIMO systems. A vital problem called beam split, i.e., the THz rainbow,  where the generated beams will split into separated physical directions at different subcarrier frequencies, has been first analyzed. We revealed that the beam split effect may cause serious array gain loss and achievable rate degradation in wideband THz massive MIMO systems. To solve this problem, we have proposed a DPP architecture by introducing a TD network into the conventional hybrid precoding architecture. By leveraging the TD network, the DPP architecture can realize a delay-phase jointly controlled beamformer, which can compensate for the array gain loss caused by the beam split effect. To realize the concept of DPP, we have further proposed a hardware structure called TTD-DPP, where the frequency-dependent phase shifts provided by the TD network are realized by a small number of TTDs. Theoretical analysis and simulation results have shown that the proposed DPP can eliminate the array gain loss caused by the beam split effect, so it can achieve more than $95$\% of the optimal array gain and achievable rate performance in wideband THz massive MIMO systems. Potential future works in this area may include other feasible implementations of DPP, improved algorithms to realize DPP, channel estimation \cite{Ref:DownT2014} and beam tracking\cite{Ref:DStrack2019} for DPP, and low-cost hardware solutions such as low-resolution PSs and low-resolution ADCs/DACs\cite{Ref:LowPS2018}. 

\section*{Appendix. Proof of Lemma 2}
\emph{Proof:} The normalized array gain achieved by the analog beamforming vector ${\mathbf{a}}_{l,m}$ on an arbitrary physical direction $\theta$ at frequency $f_{m}$ can be denoted as $\eta({\mathbf{a}}_{l,m},\theta,f_{m})=|\mathbf{f}_\mathrm{t}(2d\frac{f_{m}}{c}\theta)^{H}{\mathbf{a}}_{l,m}|$.
With $[\bar{\mathbf{a}}_{l,1}^{T},\bar{\mathbf{a}}_{l,2}^{T},\cdots,\bar{\mathbf{a}}_{l,K}^{T}]^{T}=\mathbf{f}(\theta_{l})$  and $\mathbf{p}_{l,m}=[1,e^{-j\pi\beta_{l,m}},\cdots,e^{-j\pi(K-1)\beta_{l,m}}]^{T}$, we have
\begin{equation}\label{17}
\begin{aligned}
\eta({\mathbf{a}}_{l,m},\theta,f_{m})=\frac{1}{N_\mathrm{t}}\Bigg|\sum_{k=1}^{K}\sum_{p=1}^{P}& e^{-j\pi\left[(k-1)P+(p-1)\right]\theta_{l}}e^{-j\pi(k-1)\beta_{l,m}}\\
&\times e^{j\pi\left[(k-1)P+(p-1)\right]\xi_{m}\theta}\Bigg|.
\end{aligned}
\end{equation}
By seperating the summation on $K$ and $P$, we have
\begin{equation}\label{18}
\begin{aligned}
\eta&({\mathbf{a}}_{l,m},\theta,f_{m})\\
&=\frac{1}{N_\mathrm{t}}\Bigg|\sum_{k=1}^{K}e^{-j\pi(k-1)\left[P(\theta_{l}-\xi_{m}\theta)+\beta_{l,m}\right]}\sum_{p=1}^{P}e^{-j\pi(p-1)(\theta_{l}-\xi_{m}\theta)}\Bigg|\\
&=\frac{1}{N_\mathrm{t}}|\Xi_{K}(P(\theta_{l}-\xi_{m}\theta)+\beta_{l,m})\Xi_{P}(\theta_{l}-\xi_{m}\theta)|.
\end{aligned}
\end{equation}

\begin{figure}
	\centering
	\includegraphics[width=0.44\textwidth]{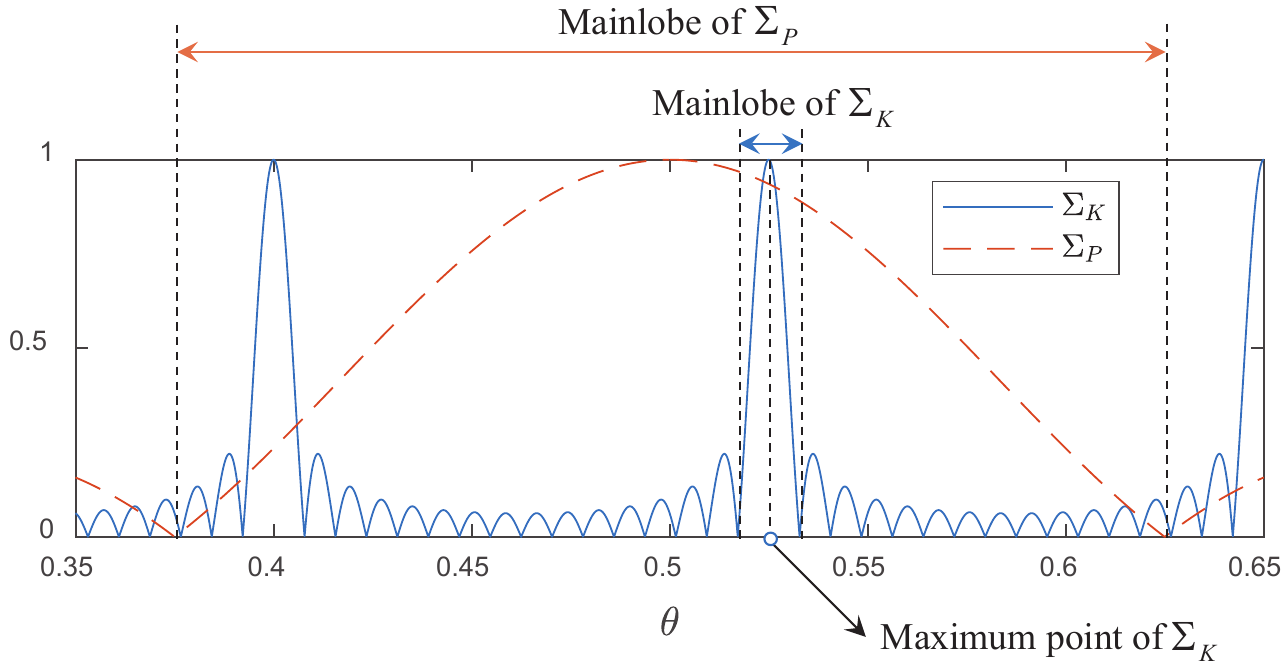}
	\caption{The two of Dirichlet sinc functions, where $\Xi_{K}$ denotes $|\frac{1}{K}\Xi_{K}(P(\theta_{l}-\xi_{m}\theta)+\beta_{l,m})|$ and $|\Xi_{P}|$ denotes $|\frac{1}{P}\Xi_{P}(\theta_{l}-\xi_{m}\theta)|$.}
\end{figure}

We can see that the array gain achieved by $\mathbf{a}_{l,m}$ is the product of two Dirichlet sinc functions as shown in Fig. 14. Due to the power-focusing property of the Dirichlet sinc function, we can analyze the array gain $\eta({\mathbf{a}}_{l,m},\theta,f_{m})$ through the mainlobes of these two Dirichlet sinc functions. For  $|\Xi_{K}(P(\theta_{l}-\xi_{m}\theta)+\beta_{l,m})|$ with respect to $\theta$, the maximum value can be achieved by setting $P(\theta_{l}-\xi_{m}\theta)+\beta_{l,m}=0$, i.e., $\theta=\theta_{K,\mathrm{max}}=\frac{\theta_{l}}{\xi_{m}}+\frac{\beta_{l,m}}{\xi_{m}P}$, and the mainlobe width of $|\Xi_{K}(P(\theta_{l,c}-\theta)+\beta_{l,m})|$ is $\frac{4}{N_{t}}$. Similarly for $|\Xi_{P}(\theta_{l,c}-\xi_{m}\theta)|$, the maximum value can be achieved when $\theta=\theta_{P,\mathrm{max}}=\frac{\theta_{l}}{\xi_{m}}$ and the mainlobe width is $\frac{4}{P}$. Considering $\beta_{l,m}\in [-1,1]$, we have $\theta_{K,\mathrm{max}}\in[\frac{\theta_{l}}{\xi_{m}}-\frac{1}{\xi_{m}P},\frac{\theta_{l}}{\xi_{m}}+\frac{1}{\xi_{m}P}],$
which means that $\theta_{K,\mathrm{max}}$ locates in the mainlobe of $|\Xi_{P}(\theta_{l}-\xi_{m}\theta)|$ whose range is $[\frac{\theta_{l}}{\xi_{m}}-\frac{2}{\xi_{m}P},\frac{\theta_{l}}{\xi_{m}}+\frac{2}{\xi_{m}P}]$. In addition,  considering $P=N_\mathrm{t}/K$, the mainlobe width of $|\Xi_{P}(\theta_{l}-\xi_{m}\theta)|$ is $\frac{4}{P}$ which is $K$ times wider than that of  $|\Xi_{K}(P(\theta_{l}-\xi_{m}\theta)+\beta_{l,m})|$, whose mainlobe width is $\frac{4}{N_\mathrm{t}}$. Therefore, we can conclude that the variation of $|\Xi_{P}(\theta_{l}-\xi_{m}\theta)|$ in the mainlobe of $|\Xi_{K}(P(\theta_{l}-\xi_{m}\theta)+\beta_{l,m})|$ is much smaller than the variation of $|\Xi_{K}(P(\theta_{l}-\xi_{m}\theta)+\beta_{l,m})|$, which is shown in Fig. 14. Therefore, the maximum value of the array gain $\eta({\mathbf{a}}_{l,m},\theta,f_{m})$ can be approximately considered to be decided by $|\Xi_{K}(P(\theta_{l}-\xi_{m}\theta)+\beta_{l,m})|$. Thus, we have
\begin{equation}\label{20}
\theta_\mathrm{opt}=\arg\max_{\theta}\eta({\mathbf{a}}_{l,m},\theta,f_{m})=\theta_{K,\mathrm{max}}=\frac{\theta_{l}}{\xi_{m}}+\frac{\beta_{l,m}}{\xi_{m}P}.
\end{equation}
Then, the array gain achieved by $\mathbf{a}_{l,m}$ at $\theta_\mathrm{opt}$ can be denoted by substituting (\ref{20}) into (\ref{18}) as
\vspace{-1mm}
\begin{equation}\label{21}
\begin{aligned}
|\eta({\mathbf{a}}_{l,m},\theta_\mathrm{opt},f_{m})|&=\frac{1}{N_{t}}|\Xi_{K}(0)\Xi_{P}(\theta_{l}-\theta_\mathrm{opt})|\\
&=\left|\frac{K}{N_{t}}\Xi_{P}(\frac{\beta_{l,m}}{P})\right|,
\end{aligned}
\end{equation}
which completes the proof.
$\hfill\blacksquare$
\vspace{-4mm}
\bibliography{Reference}

\begin{thebibliography}{10}
\providecommand{\url}[1]{#1}
\csname url@samestyle\endcsname
\providecommand{\newblock}{\relax}
\providecommand{\bibinfo}[2]{#2}
\providecommand{\BIBentrySTDinterwordspacing}{\spaceskip=0pt\relax}
\providecommand{\BIBentryALTinterwordstretchfactor}{4}
\providecommand{\BIBentryALTinterwordspacing}{\spaceskip=\fontdimen2\font plus
\BIBentryALTinterwordstretchfactor\fontdimen3\font minus
  \fontdimen4\font\relax}
\providecommand{\BIBforeignlanguage}[2]{{%
\expandafter\ifx\csname l@#1\endcsname\relax
\typeout{** WARNING: IEEEtran.bst: No hyphenation pattern has been}%
\typeout{** loaded for the language `#1'. Using the pattern for}%
\typeout{** the default language instead.}%
\else
\language=\csname l@#1\endcsname
\fi
#2}}
\providecommand{\BIBdecl}{\relax}
\BIBdecl

\bibitem{Ref:DPP2020}
J.~{Tan} and L.~{Dai}, ``Delay-phase precoding for {TH}z massive {MIMO} with
  beam split,'' in \emph{Proc. IEEE Global Commun. Conf. (GLOBECOM'19)},
  Waikoloa, USA, 2019, pp. 1--6.

\bibitem{Ref:To6G2020}
M.~{Giordani}, M.~{Polese}, M.~{Mezzavilla}, S.~{Rangan}, and M.~{Zorzi},
  ``Toward {6G} networks: Use cases and technologies,'' \emph{IEEE Commun.
  Mag.}, vol.~58, no.~3, pp. 55--61, May 2020.

\bibitem{Ref:Thz100GHz2019}
T.~S. {Rappaport}, Y.~{Xing}, O.~{Kanhere}, S.~{Ju}, A.~{Madanayake},
  S.~{Mandal}, A.~{Alkhateeb}, and G.~C. {Trichopoulos}, ``Wireless
  communications and applications above 100 {GH}z: Opportunities and challenges
  for {6G} and beyond,'' \emph{IEEE Access}, vol.~7, pp. 78\,729--78\,757, Jun.
  2019.

\bibitem{Ref:Be5GThz2018}
A.~S. {Cacciapuoti}, K.~{Sankhe}, M.~{Caleffi}, and K.~R. {Chowdhury}, ``Beyond
  {5G}: {TH}z-based medium access protocol for mobile heterogeneous networks,''
  \emph{IEEE Commun. Mag.}, vol.~56, no.~6, pp. 110--115, Jun. 2018.

\bibitem{Ref:SurThz2019}
Z.~{Chen}, X.~{Ma}, B.~{Zhang}, Y.~{Zhang}, Z.~{Niu}, N.~{Kuang}, W.~{Chen},
  L.~{Li}, and S.~{Li}, ``A survey on terahertz communications,'' \emph{China
  Commun.}, vol.~16, no.~2, pp. 1--35, Feb. 2019.

\bibitem{Ref:THzdistance2018}
D.~Li, D.~Qiao, and L.~Zhang, ``Achievable rate of indoor {TH}z communication
  systems with finite-bit {ADC}s,'' in \emph{Proc. IEEE Int. Conf. Wireless
  Common. Signal Process. (WCSP'18)}, Hangzhou, China, 2018, pp. 1--6.

\bibitem{Ref:TeraBand2014}
I.~F. {Akyildiz}, J.~M. {Jornet}, and C.~{Han}, ``Terahertz band: Next frontier
  for wireless communications,'' \emph{Phys. Commun.}, vol.~12, no.~2, pp.
  16--32, Sep. 2014.

\bibitem{Ref:Tera2011}
H.~{Song} and T.~{Nagatsuma}, ``Present and future of terahertz
  communications,'' \emph{IEEE Trans. THz Sci. Technol.}, vol.~1, no.~1, pp.
  256--263, Sep. 2011.

\bibitem{Ref:mmWaveMIMO2016}
S.~Mumtaz, J.~Rodriquez, and L.~Dai, \emph{MmWave Massive {MIMO}: A Paradigm
  for {5G}}.\hskip 1em plus 0.5em minus 0.4em\relax Academic Press, Elsevier,
  2016.

\bibitem{Ref:AoATera2018}
B.~{Peng}, K.~{Guan}, and T.~{Kurner}, ``Cooperative dynamic angle of arrival
  estimation considering space-time correlations for {T}erahertz
  communications,'' \emph{IEEE Trans.Wireless Commun.}, vol.~17, no.~9, pp.
  6029--6041, Sep. 2018.

\bibitem{Ref:BDMATera2017}
L.~{You}, X.~{Gao}, G.~Y. {Li}, X.~{Xia}, and N.~{Ma}, ``{BDMA} for
  millimeter-wave/terahertz massive {MIMO} transmission with per-beam
  synchronization,'' \emph{IEEE J. Sel. Areas Commun.}, vol.~35, no.~7, pp.
  1550--1563, Jul. 2017.

\bibitem{Ref:TrackTera2017}
X.~{Gao}, L.~{Dai}, Y.~{Zhang}, T.~{Xie}, X.~{Dai}, and Z.~{Wang}, ``Fast
  channel tracking for terahertz beamspace massive {MIMO} systems,'' \emph{IEEE
  Trans. Veh. Technol.}, vol.~66, no.~7, pp. 5689--5696, Jul. 2017.

\bibitem{Ref:HybridMili2015}
J.~A. {Zhang}, X.~{Huang}, V.~{Dyadyuk}, and Y.~J. {Guo}, ``Massive hybrid
  antenna array for millimeter-wave cellular communications,'' \emph{IEEE
  Wireless Commun.}, vol.~22, no.~1, pp. 79--87, Feb. 2015.

\bibitem{Ref:LowRF2018}
X.~{Gao}, L.~{Dai}, and A.~M. {Sayeed}, ``Low {RF}-complexity technologies to
  enable millimeter-wave {MIMO} with large antenna array for 5{G} wireless
  communications,'' \emph{IEEE Commun. Mag.}, vol.~56, no.~4, pp. 211--217,
  Apr. 2018.

\bibitem{Ref:THzChannel2019}
H.~Zhao, L.~Wei, M.~Jarrahi, and G.~J. Pottie, ``Extending spatial and temporal
  characterization of indoor wireless channels from 350 to 650 {GH}z,''
  \emph{IEEE Trans. THz Sci. Technol.}, vol.~9, no.~3, pp. 243--252, May 2019.

\bibitem{Ref:SpatiallyPre2014}
O.~E. Ayach, S.~Rajagopal, S.~Abu-Surra, Z.~Pi, and R.~W. Heath, ``Spatially
  sparse precoding in millimeter wave {MIMO} systems,'' \emph{IEEE Trans.
  Wireless Commun.}, vol.~13, no.~3, pp. 1499--1513, Mar. 2014.

\bibitem{Ref:WidebandEff2019}
K.~Wu, W.~Ni, T.~Su, R.~P. Liu, and Y.~J. Guo, ``Exploiting spatial-wideband
  effect for fast {A}o{A} estimation at lens antenna array,'' \emph{IEEE J.
  Sel. Top. Signal Process.}, vol.~13, no.~5, pp. 902--917, Sep. 2019.

\bibitem{Ref:SubWideHy2017}
S.~{Park}, A.~{Alkhateeb}, and R.~W. {Heath}, ``Dynamic subarrays for hybrid
  precoding in wideband mm{W}ave {MIMO} systems,'' \emph{IEEE Trans. Wireless
  Commun.}, vol.~16, no.~5, pp. 2907--2920, May 2017.

\bibitem{Ref:WideMIMOHy2018}
L.~{Kong}, S.~{Han}, and C.~{Yang}, ``Hybrid precoding with rate and coverage
  constraints for wideband massive {MIMO} systems,'' \emph{IEEE Trans. Wireless
  Commun.}, vol.~17, no.~7, pp. 4634--4647, Jul. 2018.

\bibitem{Ref:WideCode2016}
M.~{Cai}, K.~{Gao}, D.~{Nie}, B.~{Hochwald}, J.~N. {Laneman}, H.~{Huang}, and
  K.~{Liu}, ``Effect of wideband beam squint on codebook design in phased-array
  wireless systems,'' in \emph{Proc. IEEE Global Commun. Conf. (GLOBECOM'16)},
  Washington D. C., USA, 2016, pp. 1--6.

\bibitem{Ref:WideCode2019}
X.~{Liu} and D.~{Qiao}, ``Space-time block coding-based beamforming for beam
  squint compensation,'' \emph{IEEE Wireless Commun. Lett.}, vol.~8, no.~1, pp.
  241--244, Feb. 2019.

\bibitem{Ref:TTD2008}
H.~{Hashemi}, T.~{Chu}, and J.~{Roderick}, ``Integrated true-time-delay-based
  ultra-wideband array processing,'' \emph{IEEE Commun. Mag.}, vol.~46, no.~9,
  pp. 162--172, Sep. 2008.

\bibitem{Ref:TTD2019}
E.~Ghaderi, A.~Sivadhasan~Ramani, A.~A. Rahimi, D.~Heo, S.~Shekhar, and
  S.~Gupta, ``An integrated discrete-time delay-compensating technique for
  large-array beamformers,'' \emph{IEEE Trans. Circuits Syst. I: Reg. Papers},
  vol.~66, no.~9, pp. 3296--3306, Aug. 2019.

\bibitem{Ref:TTDTrain2020}
V.~Boljanovic, H.~Yan, E.~Ghaderi, D.~Heo, S.~Gupta, and D.~Cabric, ``Design of
  millimeter-wave single-shot beam training for true-time-delay array,'' in
  \emph{Proc. IEEE Workshop Signal Process. Adv. Wireless Commun. (SPAWC'20)},
  Atlanta, USA, 2020, pp. 1--5.

\bibitem{Ref:HybridTTD2021}
V.~Boljanovic, H.~Yan, C.-C. Lin, S.~Mohapatra, D.~Heo, S.~Gupta, and
  D.~Cabric, ``Fast beam training with true-time-delay arrays in wideband
  millimeter-wave systems,'' \emph{IEEE Trans. Circuits Syst. I: Reg. Papers},
  vol.~68, no.~4, pp. 1727--1739, Feb 2021.

\bibitem{Ref:FundWC2005}
D.~Tse and P.~Viswanath, \emph{Fundamentals of Wireless Communication}.\hskip
  1em plus 0.5em minus 0.4em\relax Cambridge, U.K.: Cambridge Univ. Press,
  2005.

\bibitem{Ref:OverMilliMIMO2016}
R.~W. {Heath}, N.~{Gonzalez-Prelcic}, S.~{Rangan}, W.~{Roh}, and A.~M.
  {Sayeed}, ``An overview of signal processing techniques for millimeter wave
  {MIMO} systems,'' \emph{IEEE J. Sel. Top. Signal Process.}, vol.~10, no.~3,
  pp. 436--453, Apr. 2016.

\bibitem{Ref:DeconMIMO2002}
A.~M. {Sayeed}, ``Deconstructing multiantenna fading channels,'' \emph{IEEE
  Trans Signal Process.}, vol.~50, no.~10, pp. 2563--2579, Oct. 2002.

\bibitem{Ref:Holo2011}
X.~{Xu}, Y.~{Pan}, P.~{Lwin}, and P.~{Liang}, ``{3D} holographic display and
  its data transmission requirement,'' in \emph{Proc. IEEE Int. Conf. Info.
  Photonics and Optical Commun. (IPOC'11)}, Jurong West, Singapore, 2011, pp.
  1--4.

\bibitem{Ref:EnHP2016}
X.~Gao, L.~Dai, S.~Han, C.~L. I, and R.~W. Heath, ``Energy-efficient hybrid
  analog and digital precoding for mmwave {MIMO} systems with large antenna
  arrays,'' \emph{IEEE J. Sel. Areas Commun.}, vol.~34, no.~4, pp. 998--1009,
  Apr. 2016.

\bibitem{Ref:PALimite2019}
J.~{Paek}, D.~{Kim}, J.~{Bang}, J.~{Baek}, J.~{Choi}, T.~{Nomiyama}, J.~{Han},
  Y.~{Choo}, Y.~{Youn}, E.~{Park}, S.~Lee, I.~{Kim}, J.~{Lee}, B.~T. {Cho}, and
  I.~{Kang}, ``15.1 {A}n 88\%-efficiency supply modulator achieving 1.08
  $\mu$s/{V} fast transition and 100 {MH}z envelope-tracking bandwidth for {5G}
  new radio {RF} power amplifier,'' in \emph{Proc. IEEE Int. Solid-State
  Circuits Conf. (ISSCC'19)}, San Francisco, USA, 2019, pp. 238--240.

\bibitem{Ref:TTD_Line2018}
M.~{Cho}, I.~{Song}, and J.~D. {Cressler}, ``A true time delay-based {SiG}e
  bi-directional {T/R} chipset for large-scale wideband timed array antennas,''
  in \emph{Proc. IEEE Radio Freq. Integr. Circuits Symp. (RFIC'18)},
  Philadelphia, USA, 2018, pp. 272--275.

\bibitem{Ref:TTD2015}
F.~Hu and K.~Mouthaan, ``A 1-20 {GH}z 400 ps true-time delay with small delay
  error in 0.13 $\mu$m {CMOS} for broadband phased array antennas,'' in
  \emph{Proc. IEEE MTT-S Int. Microw. Symp. (IMS'15)}, Phoenix, USA, 2015, pp.
  1--3.

\bibitem{Ref:NovTTD2020}
M.~H. Ghazizadeh and A.~Medi, ``Novel trombone topology for wideband
  true-time-delay implementation,'' \emph{IEEE Trans. Micro. Theory Tech.},
  vol.~68, no.~4, pp. 1542--1552, 2020.

\bibitem{Ref:TTD_Filter2017}
I.~{Mondal} and N.~{Krishnapura}, ``A 2-{GH}z bandwidth, 0.25-1.7 ns
  true-time-delay element using a variable-order all-pass filter architecture
  in 0.13 $\mu$m {CMOS},'' \emph{IEEE J. Solid-State Circuits}, vol.~52, no.~8,
  pp. 2180--2193, Aug. 2017.

\bibitem{Ref:NObeamSe2016}
X.~{Gao}, L.~{Dai}, Z.~{Chen}, Z.~{Wang}, and Z.~{Zhang}, ``Near-optimal beam
  selection for beamspace mmwave massive {MIMO} systems,'' \emph{IEEE Commun.
  Letters}, vol.~20, no.~5, pp. 1054--1057, May 2016.

\bibitem{Ref:CFAoD2018}
W.~{Shen}, L.~{Dai}, B.~{Shim}, Z.~{Wang}, and R.~W. {Heath}, ``Channel
  feedback based on {A}o{D}-adaptive subspace codebook in {FDD} massive {MIMO}
  systems,'' \emph{IEEE Trans. Commun.}, vol.~66, no.~11, pp. 5235--5248, Nov.
  2018.

\bibitem{Ref:PSorSW2016}
R.~Mendez-Rial, C.~Rusu, N.~Gonzalez-Prelcic, A.~Alkhateeb, and R.~W. Heath,
  ``Hybrid {MIMO} architectures for millimeter wave communications: Phase
  shifters or switches?'' \emph{IEEE Access}, vol.~4, pp. 247--267, Jan. 2016.

\bibitem{Ref:DownT2014}
J.~{Choi}, D.~J. {Love}, and P.~{Bidigare}, ``Downlink training techniques for
  {FDD} massive {MIMO} systems: Open-loop and closed-loop training with
  memory,'' \emph{IEEE J.Sel. Top. Signal Process.}, vol.~8, no.~5, pp.
  802--814, Oct. 2014.

\bibitem{Ref:DStrack2019}
L.~{Lian}, A.~{Liu}, and V.~K.~N. {Lau}, ``Exploiting dynamic sparsity for
  downlink {FDD}-massive {MIMO} channel tracking,'' \emph{IEEE Trans. Signal
  Process.}, vol.~67, no.~8, pp. 2007--2021, Apr. 2019.

\bibitem{Ref:LowPS2018}
P.~{Raviteja}, Y.~{Hong}, and E.~{Viterbo}, ``Millimeter wave analog
  beamforming with low resolution phase shifters for multiuser uplink,''
  \emph{IEEE Trans. Veh. Technol.}, vol.~67, no.~4, pp. 3205--3215, Apr. 2018.

\end{thebibliography}

\begin{IEEEbiography}[{\includegraphics[width=1in,height=1.25in,clip,keepaspectratio]{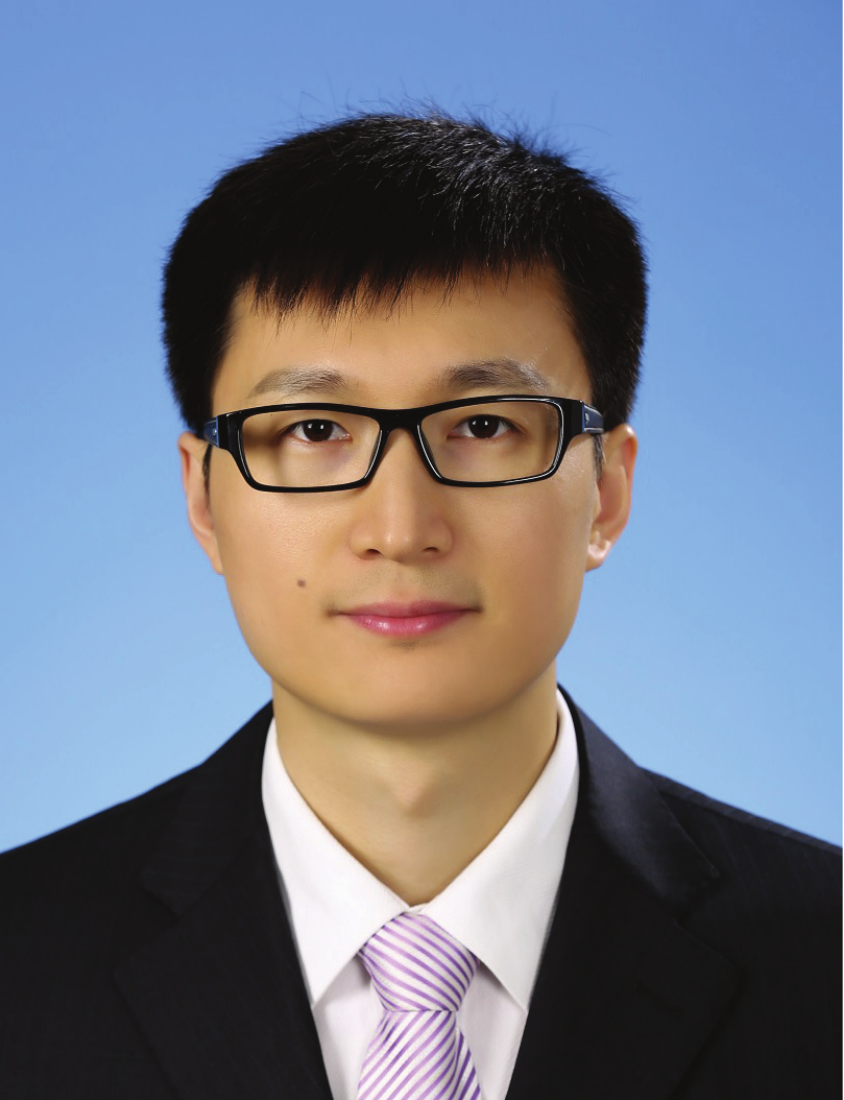}}]{Linglong Dai} (Fellow, IEEE) received the B.S. degree from Zhejiang University, Hangzhou, China, in 2003, the M.S. degree (with the highest honor) from the China Academy of Telecommunications Technology, Beijing, China, in 2006, and the Ph.D. degree (with the highest honor) from Tsinghua University, Beijing, China, in 2011. From 2011 to 2013, he was a Postdoctoral Research Fellow with the Department of Electronic Engineering, Tsinghua University, where he was an Assistant Professor from 2013 to 2016 and has been an Associate Professor since 2016. His current research interests include massive MIMO, reconfigurable intelligent surface (RIS), millimeter-wave and Terahertz communications, and machine learning for wireless communications. He has received the National Natural Science Foundation of China for Outstanding Young Scholars in 2017, the IEEE ComSoc Asia-Pacific Outstanding Young Researcher Award in 2017, and the IEEE Communications Society Leonard G. Abraham Prize in 2020. He was listed as a Highly Cited Researcher by Clarivate Analytics in 2020 and 2021. He was elevated as an IEEE Fellow in 2022.
\end{IEEEbiography}

\begin{IEEEbiography}[{\includegraphics[width=1in,height=1.25in,clip,keepaspectratio]{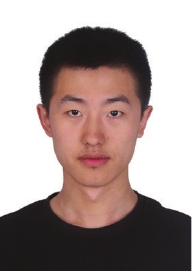}}]{Jingbo Tan} (Student Member, IEEE) received his B. S. degree in the Department of Electronic Engineering,  Tsinghua University, Beijing, China, in 2017, where he is currently pursuing his Ph. D. degree. His research interests include precoding and channel estimation in massive MIMO, THz communications, and reconfigurable intelligent surface aided systems. He has received the IEEE Communications Letters Exemplary Reviewer Award in 2018 and the Honorary Mention in the 2019 IEEE ComSoC Student Competition.
\end{IEEEbiography}

\begin{IEEEbiography}[{\includegraphics[width=1in,height=1.25in,clip,keepaspectratio]{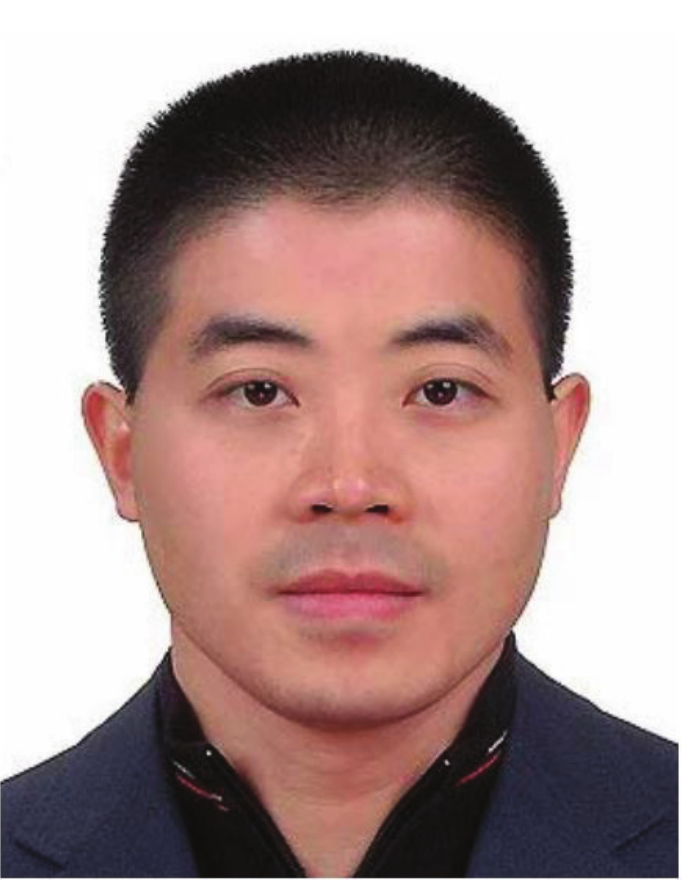}}]{Zhi Chen} (Senior Member, IEEE) received the B.Eng., M.Eng., and Ph.D. degrees in electrical engineering from the University of Electronic Science and Technology of China (UESTC) in 1997, 2000, and 2006, respectively. He joined the National Key Laboratory of Science and Technology on Communications (NCL), UESTC, in April 2006, where he has been working as a Professor since August 2013. From 2010 to 2011, he was a Visiting Scholar with the University of California at Riverside, Riverside, CA, USA. He is currently the Deputy Director of the Key Laboratory of Terahertz Technology, Ministry of Education. His current research interests include terahertz communication, 5G mobile communications, and tactile internet.
\end{IEEEbiography}

\begin{IEEEbiography}[{\includegraphics[width=1in,height=1.25in,clip,keepaspectratio]{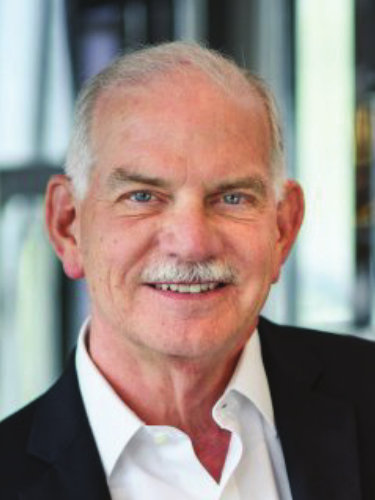}}]{H. Vincent Poor} (Life Fellow, IEEE) received the Ph.D. degree in EECS from Princeton University in 1977.  From 1977 until 1990, he was on the faculty of the University of Illinois at Urbana-Champaign. Since 1990 he has been on the faculty at Princeton, where he is currently the Michael Henry Strater University Professor. During 2006 to 2016, he served as the dean of Princeton’s School of Engineering and Applied Science. He has also held visiting appointments at several other universities, including most recently at Berkeley and Cambridge. His research interests are in the areas of information theory, machine learning and network science, and their applications in wireless networks, energy systems and related fields. Among his publications in these areas is the forthcoming book {\it Machine Learning and Wireless Communications} (Cambridge University Press). Dr. Poor is a member of the National Academy of Engineering and the National Academy of Sciences and is a foreign member of the Chinese Academy of Sciences, the Royal Society, and other national and international academies. He received the IEEE Alexander Graham Bell Medal in 2017.
\end{IEEEbiography}

\end{document}